\documentclass[11pt]{article}

\usepackage[usenames,dvipsnames]{pstricks}
\usepackage{epsfig}
\usepackage{pst-grad} % For gradients
\usepackage[space]{grffile} % For spaces in paths
\usepackage{etoolbox} % For spaces in paths
\makeatletter % For spaces in paths
\patchcmd\Gread@eps{\@inputcheck#1 }{\@inputcheck"#1"\relax}{}{}
\makeatother

\usepackage{graphicx}
\usepackage{amsmath}
\usepackage{psfrag}
\usepackage{color,soul}
\usepackage{natbib}
\usepackage{siunitx}
\usepackage{float}
\usepackage{overpic}
\usepackage{tikz}
\usepackage{multirow}
\usepackage{bm}
\usepackage{textgreek}

\usepackage{url}
\usepackage{placeins}

\usepackage{amssymb}
\usepackage{wasysym}
\usetikzlibrary{arrows}

\usepackage[noblocks]{authblk}
\usepackage[margin=1in]{geometry}
\usepackage{float}
\usepackage{enumerate}
\usepackage{caption}
\usepackage[title]{appendix}

\usepackage[usenames,dvipsnames]{pstricks}
\usepackage{epsfig}
\usepackage{pst-grad} % For gradients
\usepackage[space]{grffile} % For spaces in paths
\usepackage{etoolbox} % For spaces in paths
\makeatletter % For spaces in paths
\patchcmd\Gread@eps{\@inputcheck#1 }{\@inputcheck"#1"\relax}{}{}
\makeatother

  \DeclareTextFontCommand\textsfi{\usefont{OT1}{cmss}{m}{sl}}
  \DeclareMathAlphabet\mathsfi            {OT1}{cmss}{m}{sl}
  \DeclareTextFontCommand\textsfb{\usefont{OT1}{cmss}{bx}{n}}
  \DeclareMathAlphabet\mathsfb            {OT1}{cmss}{bx}{n}
  \DeclareTextFontCommand\textsfbi{\usefont{OT1}{cmss}{m}{sl}}
  \DeclareMathAlphabet\mathsfbi            {OT1}{cmss}{m}{sl}
  \IfFileExists{t1phv.fd}
  {\DeclareTextFontCommand\textsfbi{\usefont{T1}{phv}{b}{it}}
  \DeclareMathAlphabet\mathsfbi            {T1}{phv}{b}{it}
  }{}
  \IfFileExists{ot1phv.fd}
  {\DeclareTextFontCommand\textsfbi{\usefont{OT1}{phv}{b}{it}}
  \DeclareMathAlphabet\mathsfbi            {OT1}{phv}{b}{it}
  }{}

\newcommand\affiliation[1]{\gdef\@affiliation{\let\aff\aff@inst#1}}
\gdef\@affiliation{}

\def\email#1{Email address for correspondence: #1}
\def\aff#1{\ignorespaces\textsuperscript{#1}}
\def\corresp#1{\unskip\thanks{#1}}

\numberwithin{equation}{section}

\renewenvironment{abstract}
{\begin{quote}
\noindent \rule{\linewidth}{.5pt}\par{\bfseries \abstractname.}}
{\medskip\noindent \rule{\linewidth}{.5pt}
\end{quote}
}

\newcommand{\tc}[1]{\mathsfbi{#1}}	% Continuous tensor
\newcommand{\ii}{\mathrm{i}}

\newcommand{\cmmnt}[1]{\ignorespaces}

\definecolor{blue}{rgb}{0,0.502,1}
\definecolor{darkblue}{rgb}{0,0,0.80}
\definecolor{red}{rgb}{1,0,0}
\definecolor{green}{rgb}{0,0.8824,0}
\definecolor{gray}{rgb}{0.627,0.627,0.627}
\definecolor{black}{rgb}{0,0,0}

\newcommand\solidrule[1][1cm]{\rule[0.5ex]{#1}{.4pt}}

\newcommand{\myRuleDashDot}[1]{\textcolor{black}{\mbox{%
  \solidrule[2mm]\hspace{1mm}\solidrule[0.2mm]\hspace{1mm}\solidrule[2mm]\hspace{1mm}\solidrule[0.2mm]\hspace{1mm}\solidrule[2mm]}}}

\usepackage{hyperref}
\hypersetup{
    colorlinks=true,
    linkcolor={darkblue},
    citecolor={darkblue},
    filecolor=black,
    urlcolor=black,
    plainpages=false,
}

\title{\bf Resolvent-based estimation of space-time flow statistics}
\author[1]{\bf Aaron Towne\corresp{\email{towne@umich.edu}}}
\author[2]{\bf Adrian Lozano-Dur{\'a}n}
\author[3]{\bf Xiang Yang}

\affil[1]{\normalsize Department of Mechanical Engineering, University of Michigan, Ann Arbor, MI, USA  }
\affil[2]{\normalsize Center for Turbulence Research, Stanford University, Stanford, CA, USA}
\affil[3]{\normalsize Department of Mechanical and Nuclear Eng., Penn State University, State College, PA, USA \vspace{-1cm}}
\date{}

\begin{document}
\maketitle

% Abstract
\begin{abstract}
We develop a method to estimate space-time flow statistics from a limited set of known data.  While previous work has focused on modeling spatial or temporal statistics independently, space-time statistics carry fundamental information about the physics and coherent motions of the flow and provide a starting point for low-order modeling and flow control efforts.  The method is derived using a statistical interpretation of resolvent analysis.  The central idea of our approach is to use known data to infer the statistics of the nonlinear terms that constitute a forcing on the linearized Navier-Stokes equations, which in turn imply values for the remaining unknown flow statistics through application of the resolvent operator.  Rather than making an \textit{a priori} rank-1 assumption, our method allows the known input data to select the most relevant portions of the resolvent operator for describing the data, making it well-suited for high-rank turbulent flows.  We demonstrate the predictive capabilities of the method using two examples: the Ginzburg-Landau equation, which serves as a convenient model for a convectively unstable flow, and a turbulent channel flow at low Reynolds number. 
\end{abstract}

%% Keywords
%\begin{keywords}
%computational methods, low-dimensional models, turbulent flows
%\end{keywords}

%\note{REMAINING ISSUES}
%\begin{itemize}
%\item include adjoint resolvent stuff in appendix?
%\item should I switch to i,j,k instead of j,k,n?
%\end{itemize}

%%%%%%%%%%%%%%%%%%%%%%%%%%%%%%%%%%%%%%%%%%%%%%%%%%%%%%%%%%%%%%%%%%%%%%%%%%%%%%%
% -----------------------------------------------------------------------------
% -- Introduction
% -----------------------------------------------------------------------------

\section{Introduction}
\label{sec:intro}

Practical limitations in both experiments and simulations can lead to partial knowledge of flow statistics.  For example, an array of probes in an experiment provides information at a limited number of spatial locations and for a single flow quantity, e.g., velocity from hot-wires or pressure from microphones.  Similarly, particle image velocimetry might provide velocity data, but not thermodynamic quantities, in a limited field of view.  In simulations, one may wish to know flow statistics in a region that is not adequately resolved by the computational grid, such as unresolved near-wall regions or locations outside of the computational domain.

The ability to use available data to estimate statistics of flow quantities that are not directly accessible would be useful in each of these situations.  For example, such a method could enable recovery of full-field statistics from a discrete set of measurements, estimates of the statistics of one variable from measurement of another, or estimates for a region outside of the field of measurement or computational domain.  

Two methods for estimating unkown flow statistics from a limited set of known entries have recently been developed.  \cite{Beneddine:2016} proposed a method for estimating unknown power spectral densities (PSDs) using knowledge of the mean flow field and power spectra at a few locations.  This is accomplished using a least-squares fit at each frequency between the known power spectra and the leading singular response mode obtained from the resolvent operator \citep{Mckeon:2010}, which is derived from the linearized Navier-Stokes equations.  This strategy explicitly assumes that the spectral content at frequencies of interest is dominated by the leading resolvent mode, and the method performs well when the matching points are located in regions where this hypothesis is valid.  Specifically, excellent PSD estimates were obtained for the flow over a backward-facing step \citep{Beneddine:2016} and an initially laminar jet \citep{Beneddine:2017}.  

\cite{Zare:2017} developed a method that uses arbitrary known entries in the spatial covariance tensor to estimate the remaining unknown entries.  Their approach is also based on linearized flow equations and entails solving a convex optimization problem that determines a matrix controlling the structure and statistics of the associated nonlinear forcing terms.  The optimization problem is subject to two constraints on the estimated covariance tensor: it must reproduce the known entries and obey a Lyapunov equation that relates the forcing and flow statistics.  The constrained optimization problem is computationally demanding and requires a customized algorithm \citep{Zare:2015, Zare:2017}.  

The objective of the present paper is to build on these previous methods to estimate unknown two-point space-time flow statistics.  Both the PSDs (one-point temporal statistics) and spatial covariances (two-point spatial statistics) are subsets of two-point space-time correlations, so our approach represents a generalization of these previous methods.  This is an important step since two-point space-time statistics contain additional, fundamental information about the flow.  In particular, they carry information about coherent motions within the flow, and can even be used to define the concept of a coherent structure \citep{Towne2018spectral}.  Moreover, space-time correlations can be used to obtain real-time estimates of the flow state via convolution with a time-varying input signal \citep{Sasaki2017Real}.  

\sloppypar{The method developed in this paper borrows ideas from both of the previously mentioned methods.  Like \cite{Beneddine:2016}, our method is built upon the resovent formalism of \cite{Mckeon:2010}. The resolvent operator is derived from the Navier-Stokes equations linearized about the turbulent mean flow and constitutes a transfer function in the frequency domain between terms that are nonlinear and linear with respect to fluctuations to the mean.  Whereas \cite{Beneddine:2016} constructed their model using only the first singular mode of the resolvent operator (obtained via singular value decomposition), our model relaxes this \textit{a priori} assumption and allows the known data to self-select the relevent portion of the resolvent operator.  This makes the method more applicable to turbulent flows, in which the leading resolvent mode may account for only a modest fraction of the total flow energy \citep{Schmidt2018spectral}, and allows us to extend the method to cross-spectra in addition to power spectra.}  

Our approach also follows the underlying strategy employed by \cite{Zare:2017} of using the known data to infer the statistics of the unknown nonlinear terms that act as a forcing on the linearized equations.  Their method assumes the nonlinear terms to have the same spatial correlation at all frequencies, in contrast to recent findings \citep{Rosenberg:2016,Towne:2017a}.  Our method relaxes this assumption and allows different spatial correlations for each frequency.  Moreover, our frequency-domain formulation is algorithmically simple, requiring only basic linear algebra manipulations and avoiding Lyapunov equations and the need for external optimization routines.  

The objective and capabilities of our method are fundamentally different from those of the classical method of linear stochastic estimation \citep{Adrian:1994, Bonnet1994} and related approaches that have recently been investigated \citep[e.g.,][]{Encinar:2018}.  In these methods, cross-correlations between input quantities and output quantities of interest must be known \textit{a priori} and are used to estimate instantaneous values or conditional averages for the quantities of interest.  In contrast, our method assumes no knowledge of the output statistics (or its cross correlation with input quantities), and instead uses input data alone to estimate space-time statistics of the output quantities of interest.   Accordingly, our method could in fact be used to obtain an estimate of the statistics required to perform linear stochastic estimation.  

The remainder of this paper is organized as follows.  The method is derived and described in \S~\ref{Sec:method} and demonstrated in \S~\ref{Sec:results} using two examples: a simple model problem given by the Ginzburg-Landau equation and a turbulent channel flow.  Finally,  \S~\ref{Sec:conclusions} summarizes the paper and discusses further improvements and applications of the method.

%%%%%%%%%%%%%%%%%%%%%%%%%%%%%%%%%%%%%%%%%%%%%%%%%%%%%%%%%%%%%%%%%%%%%%%%%%%%%%%
% -----------------------------------------------------------------------------
% -- SECTION: 
% -----------------------------------------------------------------------------
\section{Method}
\label{Sec:method}

Our method for estimating space-time flow statistics from limited measurements is developed in this section.  After precisely defining the objective, we develop our approach to the problem and provide some alternative interpretations of the method, which help to elucidate its properties.  

\subsection{Objective}

Consider a state vector of flow variables $\boldsymbol{q}(\boldsymbol{x},t)$ that describe a flow, e.g., velocities and thermodynamic variables.  The independent variables $\boldsymbol{x}$ and $t$ represent the spatial dimensions of the problem and time, respectively. Now suppose that the two-point space-time statistics are known for a reduced set of variables
\begin{equation}
\label{Eq:y_eq_Cq}
\boldsymbol{y} = \mathcal{C}\boldsymbol{q},
\end{equation}
where the linear operator $\mathcal{C}(\boldsymbol{x})$ selects any desired subset or linear combination of $\boldsymbol{q}$.  The problem objective can now be precisely stated in terms of two-point space-time correlation tensors:
\begin{subequations}
\label{Eq:Cuu_def}
\begin{alignat}{1}
\mathrm{given} \qquad \tc{C}_{\boldsymbol{y}\boldsymbol{y}}(\boldsymbol{x},\boldsymbol{x}^{\prime},\tau) =& \, E \left\{ \boldsymbol{y}(\boldsymbol{x},t) \boldsymbol{y}^{*}(\boldsymbol{x}^{\prime},t+\tau) \right\}, \\
\mathrm{estimate} \qquad \tc{C}_{\boldsymbol{q}\boldsymbol{q}}(\boldsymbol{x},\boldsymbol{x}^{\prime},\tau) =& \, E \left\{ \boldsymbol{q}(\boldsymbol{x},t) \boldsymbol{q}^{*}(\boldsymbol{x}^{\prime},t+\tau) \right\}.
\end{alignat}
\end{subequations}
Here, $E\left\{\cdot\right\}$ is the expectation operator over time and the asterisk superscript indicates a Hermitian transpose.  

Using the relationship between space-time correlation tensors and the cross-spectral density (CSD) tensors
\begin{equation}
\label{Eq:COR2CSD}
\tc{S}(\boldsymbol{x},\boldsymbol{x}^{\prime},\omega) = \int \limits_{-\infty}^{\infty} \tc{C}(\boldsymbol{x},\boldsymbol{x}^{\prime},\tau) e^{i \omega \tau} d \tau,
\end{equation}
this objective can be equivalently stated in the frequency domain for statistically stationary flows: 
\begin{subequations}
\label{Eq:Suu_def}
\begin{alignat}{1}
\mathrm{given} \qquad \tc{S}_{\boldsymbol{y}\boldsymbol{y}}(\boldsymbol{x},\boldsymbol{x}^{\prime},\omega) =& \, E \left\{ \hat{\boldsymbol{y}}(\boldsymbol{x},\omega) \hat{\boldsymbol{y}}^{*}(\boldsymbol{x}^{\prime},\omega) \right\}, \label{Eq:Syy_def} \\
\mathrm{estimate} \qquad \tc{S}_{\boldsymbol{q}\boldsymbol{q}}(\boldsymbol{x},\boldsymbol{x}^{\prime},\omega) =& \, E \left\{ \hat{\boldsymbol{q}}(\boldsymbol{x},\omega) \hat{\boldsymbol{q}}^{*}(\boldsymbol{x}^{\prime},\omega) \right\}, \label{Eq:Sqq_def}
\end{alignat}
\end{subequations}
where $\hat{\boldsymbol{y}}(\boldsymbol{x},\omega)$ and $\hat{\boldsymbol{q}}(\boldsymbol{x},\omega)$ are the temporal Fourier transforms of $\boldsymbol{y}$ and $\boldsymbol{q}$, respectively, and the expectation is now taken over realizations of the flow \citep{Bendat:1990}.

\subsection{Approach}

Our approach to this problem relies on the resolvent operator obtained from the linearized flow equations and its connection with the remaining nonlinear terms \citep{Mckeon:2010}. Begin with nonlinear flow equations of the form
\begin{equation}
\label{Eq:intro_model_F}
{\mathcal{G}} \frac{\partial \boldsymbol{q}}{\partial t} = \boldsymbol{\mathcal{F}}\left( \boldsymbol{q} \right),
\end{equation}
where ${\mathcal{G}}$ and $\boldsymbol{\mathcal{F}}$ are linear and nonlinear operators, respectively.  Both compressible and incompressible Navier-Stokes equations can be cast in this form, and ${\mathcal{G}}$ is singular in the incompressible case to account for algebraic divergence-free condition.  Alternatively, the incompressible equations can be written with a non-singular ${\mathcal{G}}$ by projecting into a divergence-free basis to eliminate the continuity equation \citep{Meseguer:2003}.  Additional transport equations can also be included.

Applying the Reynolds decomposition
\begin{equation}
\label{Eq:intro_model_reynolds}
\boldsymbol{q}\left(\boldsymbol{x},t \right) = \bar{\boldsymbol{q}}\left(\boldsymbol{x}\right) + \boldsymbol{q}^{\prime}\left(\boldsymbol{x},t \right),
\end{equation}
where $\bar{\boldsymbol{q}}\left(\boldsymbol{x}\right)$ is the mean (time-averaged) flow, to (\ref{Eq:intro_model_F}) and isolating the terms that are linear in $\boldsymbol{q}^{\prime}$ yields an equation of the form
\begin{equation}
\label{Eq:intro_model_LNS_time}
{\mathcal{G}} \frac{\partial \boldsymbol{q}^{\prime}}{\partial t} - \mathcal{A}\left(\bar{\boldsymbol{q}}\right) \boldsymbol{q}^{\prime} = \boldsymbol{f}\left(\bar{\boldsymbol{q}},\boldsymbol{q}^{\prime}\right),
\end{equation}
where 
\begin{equation}
\label{Eq:intro_model_A}
\mathcal{A}\left(\bar{\boldsymbol{q}}\right) = \frac{\partial \boldsymbol{\mathcal{F}}}{\partial \boldsymbol{q}}\left(\bar{\boldsymbol{q}}\right)
\end{equation}
is the linearized Navier-Stokes operator and $\boldsymbol{f}$ contains the remaining nonlinear terms.  Similarly, (\ref{Eq:y_eq_Cq}) becomes
\begin{equation}
\label{Eq:y_eq_Cq_prime}
\boldsymbol{y}^{\prime} = \mathcal{C}\boldsymbol{q}^{\prime}.
\end{equation}

In the frequency domain, (\ref{Eq:intro_model_LNS_time}) and~(\ref{Eq:y_eq_Cq_prime}) can be manipulated to give
\begin{subequations}
\label{Eq:u_hat_eq_Ru_f}
\begin{align}
\hat{\boldsymbol{y}} =& \mathcal{R}_{y} \hat{\boldsymbol{f}},\\
\hat{\boldsymbol{q}} =& \mathcal{R}_{q} \hat{\boldsymbol{f}},
\end{align}
\end{subequations}
where 
\begin{subequations}
\begin{align}
\mathcal{R}_{\boldsymbol{y}}(\boldsymbol{x},\omega) =& \, \mathcal{C} \left(i \omega \mathcal{G} - \mathcal{A} \right)^{-1}, \\
\mathcal{R}_{\boldsymbol{q}}(\boldsymbol{x},\omega) =& \,\left(i \omega \mathcal{G} - \mathcal{A} \right)^{-1}
\end{align}
\end{subequations}
are resolvent operators associated with $\hat{\boldsymbol{y}}$ and $\hat{\boldsymbol{q}}$, respectively.
 
Using (\ref{Eq:Suu_def}) and~(\ref{Eq:u_hat_eq_Ru_f}), the CSD tensors can be written in terms of these resolvent operators as
\begin{subequations}
\label{Eq:Suu_res}
\begin{align}
\tc{S}_{\boldsymbol{y}\boldsymbol{y}} =& \, \mathcal{R}_{\boldsymbol{y}} \tc{S}_{\boldsymbol{f}\boldsymbol{f}} \mathcal{R}^{*}_{\boldsymbol{y}},\label{Eq:Syy_res} \\
\tc{S}_{\boldsymbol{q}\boldsymbol{q}} =& \, \mathcal{R}_{\boldsymbol{q}} \tc{S}_{\boldsymbol{f}\boldsymbol{f}} \mathcal{R}^{*}_{\boldsymbol{q}}, \label{Eq:Sqq_res}
\end{align}
\end{subequations}
where $\tc{S}_{\boldsymbol{f}\boldsymbol{f}}(\boldsymbol{x},\boldsymbol{x}^{\prime},\omega) = E \{ \hat{\boldsymbol{f}}(\boldsymbol{x},\omega) \hat{\boldsymbol{f}}^{*}(\boldsymbol{x}^{\prime},\omega) \}$ is the CSD tensor of the nonlinear term $\boldsymbol{f}$ \citep{Semeraro:2016, Towne:2016b, Towne2018spectral}.  We emphasize that no approximation has been made to this point; (\ref{Eq:Suu_res}) is an exact expression of the Navier-Stokes equations.

To obtain an approximation of the desired statistics $\tc{S}_{\boldsymbol{q}\boldsymbol{q}}$, we use the known statistics $\tc{S}_{\boldsymbol{y}\boldsymbol{y}}$ to estimate $\tc{S}_{\boldsymbol{f}\boldsymbol{f}}$.  The salient question then becomes: how much can we learn about $\tc{S}_{\boldsymbol{f}\boldsymbol{f}}$ from $\tc{S}_{\boldsymbol{y}\boldsymbol{y}}$?  An answer is provided by examining the singular value decomposition (SVD)
\begin{subequations}
\label{Eq:Ry_SVD}
\begin{align}
\mathcal{R}_{\boldsymbol{y}} =& \, \boldsymbol{U}_{\boldsymbol{y}} \boldsymbol{\Sigma}_{\boldsymbol{y}} \boldsymbol{V}_{\boldsymbol{y}}^{*} \label{Eq:Ry_SVD_1} \\
=& \, \boldsymbol{U}_{\boldsymbol{y}} \left[ \begin{array}{cc} \boldsymbol{\Sigma}_{1} & \boldsymbol{0}\end{array} \right] \left[ \boldsymbol{V}_{1} \,\,\,\boldsymbol{V}_{2} \right]^{*}. \label{Eq:Ry_SVD_2}
\end{align}
\end{subequations}
The columns of the matrices $\boldsymbol{V}_{\boldsymbol{y}}$ and $\boldsymbol{U}_{\boldsymbol{y}}$ correspond to input and output modes that form orthonormal bases for $\hat{\boldsymbol{f}}$ and $\hat{\boldsymbol{y}}$, respectively.  The rectangular matrix $\boldsymbol{\Sigma}_{\boldsymbol{y}}$ determines the gain of each of the input modes to the output.  Since the rank of $\mathcal{R}_{\boldsymbol{y}}$ can be no greater than the number of entries in $\boldsymbol{y}$, i.e., the number of locations/quantities for which the statistics are known, many of the input modes have no impact on the output.  Accordingly, the SVD can be written in the form of (\ref{Eq:Ry_SVD_2}), where the diagonal $\boldsymbol{\Sigma}_{1}$ contains the non-zero singular values and the blocks $\boldsymbol{V}_{1}$ and $\boldsymbol{V}_{2}$ contain input modes that have non-zero and zero gain, respectively.  It is important to note that these resolvent modes are different from those usually studied, which are given by the SVD of $\mathcal{R}_{\boldsymbol{q}}$ \citep[e.g.,][]{Mckeon:2010,Schmidt2018spectral}.  

The distinction between input modes that do or do not impact the output can be used to isolate the part of $\tc{S}_{\boldsymbol{f}\boldsymbol{f}}$ that can be educed from knowledge of $\tc{S}_{\boldsymbol{y}\boldsymbol{y}}$.  Since $\boldsymbol{V}_{\boldsymbol{y}}$ provides a complete basis for $\hat{\boldsymbol{f}}$, $\tc{S}_{\boldsymbol{f}\boldsymbol{f}}$ can be expanded as
\begin{equation}
\tc{S}_{\boldsymbol{f}\boldsymbol{f}} = \left[ \boldsymbol{V}_{1} \,\,\,\boldsymbol{V}_{2} \right]     \left[ \begin{array}{cc} \boldsymbol{E}_{11} & \boldsymbol{E}_{12} \\ \boldsymbol{E}_{21} & \boldsymbol{E}_{22} \end{array} \right]\left[ \boldsymbol{V}_{1} \,\,\,\boldsymbol{V}_{2} \right]^{*},
\end{equation}
where the matrices $\boldsymbol{E}_{ij}$ represent correlation between expansion coefficients associated with each input mode \citep[see][]{Towne2018spectral}.  Inserting this expansion into (\ref{Eq:Syy_res}) and using (\ref{Eq:Ry_SVD_2}) to simplify the expression gives rise to the equation
\begin{equation}
\label{Eq:Syy_E11}
\tc{S}_{\boldsymbol{y}\boldsymbol{y}} = \boldsymbol{U}_{\boldsymbol{y}} \boldsymbol{\Sigma}_{1} \boldsymbol{E}_{11}\boldsymbol{\Sigma}_{1} \boldsymbol{U}_{\boldsymbol{y}}^{*}.
\end{equation}
This means that only the part of $\tc{S}_{\boldsymbol{f}\boldsymbol{f}}$ associated with $\boldsymbol{E}_{11}$ impacts the observed statistics $\tc{S}_{\boldsymbol{y}\boldsymbol{y}}$; the remaining $\boldsymbol{E}_{ij}$ terms have no impact and are thus unobservable from these known data.  Consequently, $\boldsymbol{E}_{11}$ contains all of the information about $\tc{S}_{\boldsymbol{f}\boldsymbol{f}}$ that can be inferred from $\tc{S}_{\boldsymbol{y}\boldsymbol{y}}$.  Using the orthonormality of $\boldsymbol{U}_{\boldsymbol{y}}$, (\ref{Eq:Syy_E11}) gives
\begin{equation}
\label{Eq:E11_eq}
\boldsymbol{E}_{11} = \boldsymbol{\Sigma}_{1}^{-1}\boldsymbol{U}_{\boldsymbol{y}}^{*}  \tc{S}_{\boldsymbol{y}\boldsymbol{y}} \boldsymbol{U}_{\boldsymbol{y}} \boldsymbol{\Sigma}_{1}^{-1}.
\end{equation}

The remaining terms $\boldsymbol{E}_{22}$ and $\boldsymbol{E}_{12} = \boldsymbol{E}_{21}^{*}$ (this equality is required to make $\tc{S}_{\boldsymbol{f}\boldsymbol{f}}$ Hermetian) can be arbitrarily chosen without impacting $\tc{S}_{\boldsymbol{y}\boldsymbol{y}}$, but these terms will impact $\tc{S}_{\boldsymbol{q}\boldsymbol{q}}$ and therefore must be modeled.  The simplest choice, and the one used in the remainder of this paper, is to set these unknown terms to zero, leading to the approximation
\begin{equation}
\label{Eq:Sff_approx}
\tc{S}_{\boldsymbol{f}\boldsymbol{f}} \approx \left[ \boldsymbol{V}_{1} \,\,\,\boldsymbol{V}_{2} \right]     \left[ \begin{array}{cc} \boldsymbol{E}_{11} & 0 \\ 0 & 0 \end{array} \right]\left[ \boldsymbol{V}_{1} \,\,\,\boldsymbol{V}_{2} \right]^{*} = \boldsymbol{V}_{1} \boldsymbol{E}_{11} \boldsymbol{V}_{1}^{*}.
\end{equation}
We show in Appendix~\ref{Sec:app_LS_forcing} that this choice is identical to taking the least-squares approximation of $\tc{S}_{\boldsymbol{f} \boldsymbol{f}}$, which can be obtained by applying the pseudo-inverse of $\mathcal{R}_{y}$ and its complex conjugate to the left and right sides of (\ref{Eq:Syy_res}), respectively.  Therefore, this approximation corresponds to choosing the smallest forcing (in an appropriate norm) that reproduces the known flow statistics. 

Inserting (\ref{Eq:Sff_approx}) into (\ref{Eq:Sqq_res}) gives the corresponding approximation of the desired flow statistics
\begin{equation}
\label{Eq:Sqq_approx}
\tc{S}_{\boldsymbol{q}\boldsymbol{q}} \approx \mathcal{R}_{\boldsymbol{q}} \boldsymbol{V}_{1} \boldsymbol{E}_{11} \boldsymbol{V}_{1}^{*} \mathcal{R}^{*}_{\boldsymbol{q}}.
\end{equation}
By construction, the known statistics used as input are exactly recovered, ensuring that the approximation converges in the limit of full knowledge of the flow statistics.  Other approximations can be obtained by choosing the unknown $\boldsymbol{E}_{ij}$ terms differently; a few possibilities are discussed in \S~\ref{Sec:conclusions}.  The estimated space-time correlation tensor can be recovered from the estimated CSD via the inverse Fourier transform
\begin{equation}
\label{Eq:CSD2COR}
\tc{C}(\boldsymbol{x},\boldsymbol{x}^{\prime},\tau) = \int \limits_{-\infty}^{\infty} \tc{S}(\boldsymbol{x},\boldsymbol{x}^{\prime},\omega) e^{-i \omega \tau} d \omega,
\end{equation}

The method can also be understood in terms of a resolvent-mode expansion of $\tc{S}_{\boldsymbol{q}\boldsymbol{q}}$.  The standard resolvent modes associated with the linearized flow equations are defined by the SVD $\mathcal{R}_{\boldsymbol{q}} = \, \boldsymbol{U}_{\boldsymbol{q}} \boldsymbol{\Sigma}_{\boldsymbol{q}} \boldsymbol{V}_{\boldsymbol{q}}^{*}$.  Equation~(\ref{Eq:Sqq_approx}) can then be written as
\begin{equation}
\label{Eq:Sqq_approx2}
\tc{S}_{\boldsymbol{q}\boldsymbol{q}} \approx \tc{U}_{\boldsymbol{q}} \boldsymbol{\Sigma}_{\boldsymbol{q}} \tc{S}_{\beta\beta} \boldsymbol{\Sigma}_{\boldsymbol{q}} \tc{U}^{*}_{\boldsymbol{q}},
\end{equation}
where
\begin{equation}
\tc{S}_{\beta\beta} = \boldsymbol{V}_{\boldsymbol{q}}^{*} \boldsymbol{V}_{1} \boldsymbol{E}_{11} \boldsymbol{V}_{1}^{*} \boldsymbol{V}_{\boldsymbol{q}}
\end{equation} 
is the CSD of the expansion coefficients in a resolvent-mode expansion of $\hat{\boldsymbol{q}}$ \citep{Towne2018spectral}.  In general, $\tc{S}_{\beta\beta}$ can project onto any of the resolvent output modes in $\tc{U}_{\boldsymbol{q}}$.  Thus, the known statistics $\tc{S}_{\boldsymbol{y}\boldsymbol{y}}$, through their influence on $\boldsymbol{E}_{11}$, determine which resolvent modes participate in the estimate of $\tc{S}_{\boldsymbol{q}\boldsymbol{q}}$.  This can be contrasted with the rank-1 model described earlier, in which only the leading resolvent mode is allowed to contribute.

%%%%%%%%%%%%%%%%%%%%%%%%%%%%%%%%%%%%%%%%%%%%%%%%%%%%%%%%%%%%%%%%%%%%%%%%%%%%%%%
% -----------------------------------------------------------------------------
% -- SECTION: Example results
% -----------------------------------------------------------------------------
\section{Examples}
\label{Sec:results}

In this section, our method is demonstrated and analyzed using two example problems: the complex Ginzburg-Landau equation and a turbulent channel flow.

% -----------------------------------------------------------------------------
% -- SUBSECTION: GL
% -----------------------------------------------------------------------------
\subsection{Ginzburg-Landau equation} 
\label{Sec:GL}

The Ginzburg-Landau equation has been used by several previous authors \citep[e.g.,][]{Hunt:1991,Bagheri:2009, Chen:2011, Towne2018spectral} as a convenient one-dimensional model that mimics key properties of the linearized Navier-Stokes operator for real flows, such as a turbulent jet \citep{Schmidt2018spectral}.  The linearized operator takes the form
\begin{equation}
\label{Eq:exGL:A}
\mathcal{A} = -\nu \frac{\partial}{\partial x} + \gamma \frac{\partial^{2}}{\partial x^{2}} + \mu(x).
\end{equation}
Several variants of the function  $\mu(x)$ have been used in the literature; here the quadratic form
\begin{equation}
\label{Eq:exGL_mu}
\mu(x) = (\mu_{0} - c_{\mu}^2) + \frac{\mu_{2}}{2} x^{2}
\end{equation}
is adopted  \citep{Hunt:1991,Bagheri:2009, Chen:2011}.  All of the parameters in equations~(\ref{Eq:exGL:A}) - (\ref{Eq:exGL_mu}) are set to the values used by \cite{Towne2018spectral}.  With these parameters, the leading singular value of $\mathcal{R}_{\boldsymbol{q}}$ at its peak frequency is 10 times larger than the second singular value, which is a typical value for real flows.  
Following \cite{Bagheri:2009}, the equations are discretized with a pseudo-spectral approach using $N=220$ Hermite polynomials.

The discretized equations are stochastically excited in the time domain using forcing terms with prescribed statistics identical to those used by \citep{Towne2018spectral}.  In particular, the forcing is generated by convolving band-limited white noise with a kernel of the form
\begin{equation}
\label{Eq:ex:filter}
g(x,x^{\prime}) = \frac{1}{\sqrt{2\pi} \sigma_{f}} \exp \left[ -\frac{1}{2} \left( \frac{x-x^{\prime}}{\sigma_{f}} \right)^{2} \right] \exp \left[ \ii 2\pi \frac{x - x^{\prime}}{\lambda_{f}} \right],
\end{equation}
where $\sigma_{f}$ is the standard deviation of the envelope and $\lambda_{f}$ is the wavelength of the filter.  This leads to a forcing that is white-in-time up to the cut-off frequency but that has non-zero spatial correlation in the form of~(\ref{Eq:ex:filter}) but with $\sigma_{f}$ replaced with $\sqrt{2}\sigma_{f}$.  This form of the forcing statistics is qualitatively similar to those of the nonlinear forcing terms in real flows, such as a turbulent jet \citep{Towne:2017a}.  We use $\sigma_{f} = 4$ and $\lambda_{f} = 20$.  

Although these forcing statistics are prescribed in this model problem and therefore known, this knowledge is not made available to the estimation procedure. The equations are integrated using a fourth-order embedded Runge-Kutta method \citep{Shampine:1997}, and a total of $10000$ snapshots of the solution are collected with spacing $\Delta t = 0.5$, leading to a Nyquist frequency of $\omega_{Nyquist} = 2\pi$.  The CSD of the solution is computed from these data using Welch's (\citeyear{Welch:1967}) method.  

For the majority of the following analysis, $\boldsymbol{y}$ is defined to correspond to data obtained from three probes located at $x = -10$, 0 and 10.  Other choices are considered in \S~\ref{Sec:GL_compare}.

%=========================================================================
% -- Power spectra -------------------------------------------------------
\subsubsection{Power spectra}
\label{Sec:GL_PSD}

\begin{figure}
\input{PSD_estimate_3p_JFM.tex}
\centering
\includegraphics[trim=0cm 0.0625in 0cm 0.0cm, clip=true,width=6.25in]{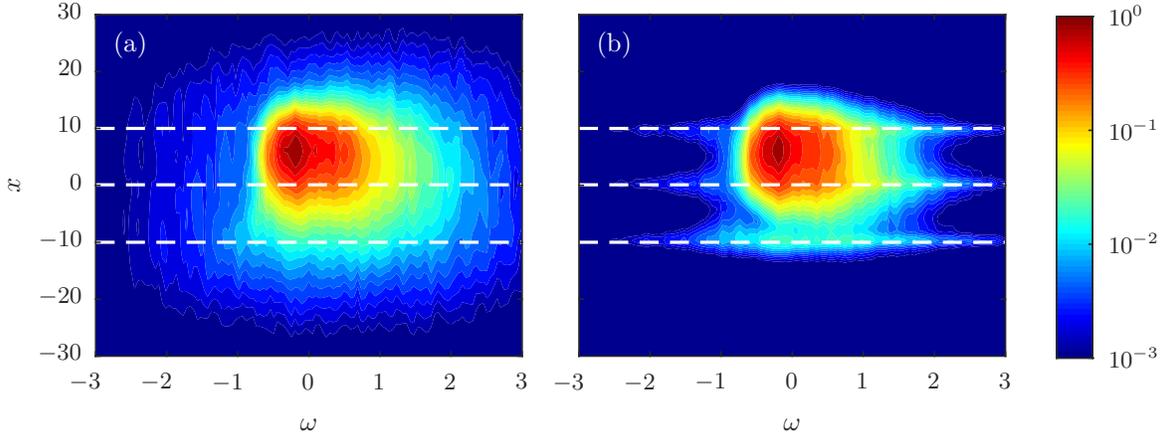}
\caption{Power spectral density (PSD) as a function of $\omega$ and $x$ for the Ginzburg-Landau model problem: (a) true PSD and (b) estimated PSD using three probes at the locations of the dashed lines.}
\label{fig:PSD_estimate_3p}
\end{figure}

The PSD is contained in the diagonal entries of $\tc{S}_{\boldsymbol{q}\boldsymbol{q}}$.  The true power-spectral density for the Ginzburg-Landau model problem is shown as a function of $\omega$ and $x$ in figure~\ref{fig:PSD_estimate_3p}(a).  A single peak is observed at $\omega \approx -0.2$ and $x \approx 5$, and the amplitude remains above 1\% of the peak over a range of about $-0.75 < \omega < 2$ and $-20<x<15$.  The dashed lines show the $x$ locations where the data is taken as known, and the estimation procedure will attempt to reconstruct the PSD elsewhere.    

The approximation of the PSD obtained using these three probes is shown in figure~\ref{fig:PSD_estimate_3p}(b).  By construction, the approximation is exact at the probe locations.  The peak is well captured and the agreement is good in high-energy regions.  In the lower-energy regions, the PSD is under-predicted away from the probe locations.  This is a consequence of neglecting the undetermined portions of the forcing.  It is likely that additional improvements could be obtained by modeling these undetermined portions of the forcing, as discussed in \S~\ref{Sec:conclusions}.

% which can be shown to always lead to global ($x$-integrated) PSD estimates that are less than or equal to the true value at each frequency

%=========================================================================
% -- Cross-spectra -------------------------------------------------------
\subsubsection{Cross-spectra}
\label{Sec:GL_CSD}

\begin{figure}
\input{CSD_estimate_3p_JFM.tex}
\centering
\includegraphics[trim=0cm 0.0625in 0cm 0.0cm, clip=true,width=6.5in]{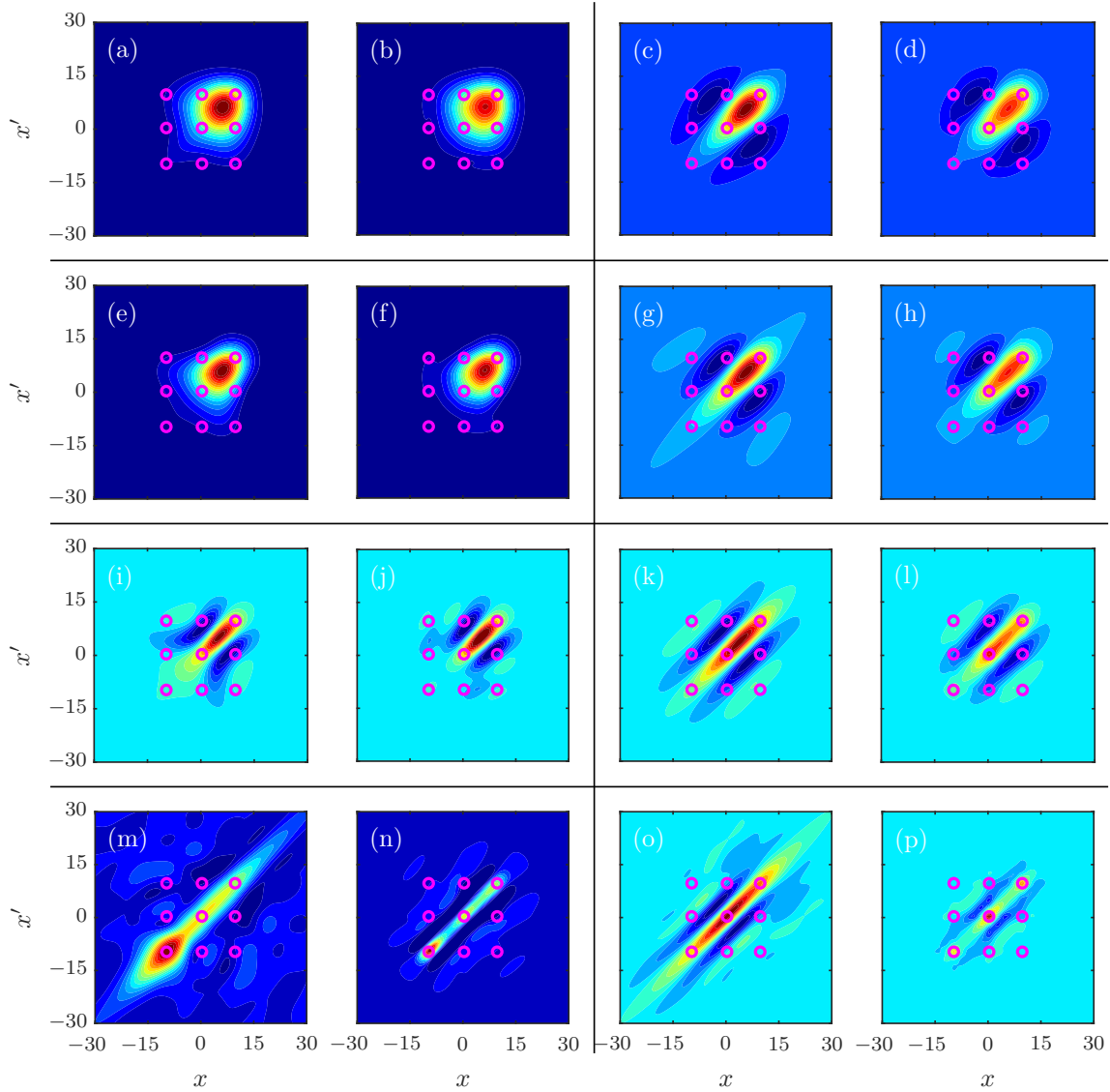}
\caption{Cross-spectral density (CSD) for the frequencies (a-b) $\omega = 0$; (e-f) $\omega = -0.2$; (i-j) $\omega = -0.6$; (m-n) $\omega = -1$; (c-d) $\omega = 0.4$; (g-h) $\omega = 0.6$; (k-l) $\omega = 1$; (o-p) $\omega = 2$.  In each case, the left-hand plot shows the true CSD and the right-hand plot shows the estimated values using three probes, which lead to known CSD values at the locations indicated by the small circles. The contour levels are the same for the true and estimated data at each frequency and range from the minimum to maximum values of the true CSD. }
\label{fig:CSD_estimate_3p}
\end{figure}

The CSD estimates are evaluated next.  Figure~\ref{fig:CSD_estimate_3p} compares the real part of the true and modeled CSD at eight frequencies, which are listed in the caption.  The contour levels are the same for the true and estimated data at each frequency and range from the minimum to maximum values of the true CSD.  The circles indicate the locations where the CSD is known and the remaining values are to be estimated.  The first six frequencies (panels a-l) fall within the high-energy region observed in figure~\ref{fig:PSD_estimate_3p}.  In these cases, the estimates are accurate and track the length scales and shape of the CSD as a function of frequency.  The final two frequencies (panels m-p) fall in low-energy regions.  The basic trends in the length scales and shape are still captured, but the estimates are not quantitatively accurate.

%=========================================================================
% -- Space-time correlations----------------------------------------------

\subsubsection{Space-time correlations}
\label{Sec:GL_COR}

\begin{figure}
\input{COR_estimate_Lines_JFM.tex}
\centering
\includegraphics[trim=0cm 0.0625in 0cm 0.0cm, clip=true,width=6.5in]{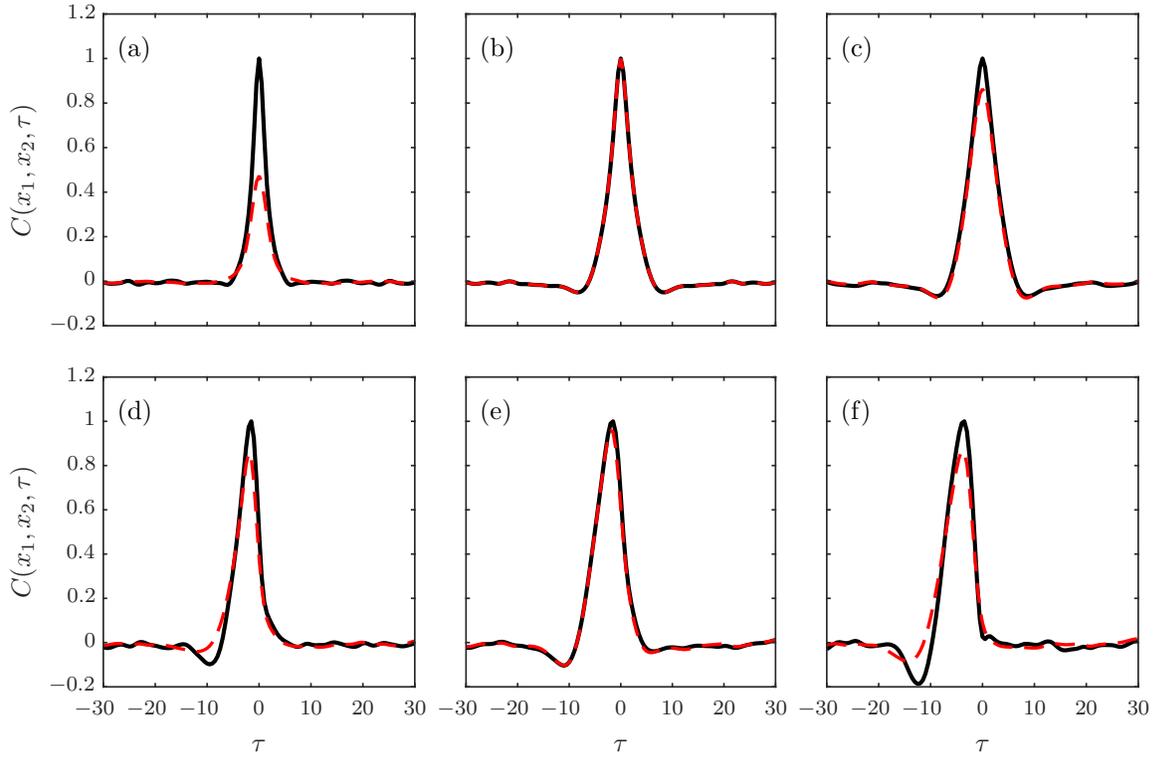}
\caption{Cross-correlation as a function of time lag $\tau$ for (a) $x_{1} = x_{2} = -5$;(b) $x_{1} = x_{2} = 0$; (c) $x_{1} = x_{2} = 5$; (d) $x_{1}=-5$,$ x_{2} = 0$; (e) $x_{1}=0$,$ x_{2} = 5$; (d) $x_{1}=-5$,$ x_{2} = 5$.  The solid lines show the true values, and the dashed lines show the estimates values using three probes at $x = -10,0$ and 10. Both the true and estimated curves in each plot have been scaled by the maximum value of the true correlation. }
\label{fig:COR_estimate_Lines}
\end{figure}

\begin{figure}
\input{COR_estimate_3p_JFM.tex}
\centering
\includegraphics[trim=0cm 0.0625in 0cm 0.0cm, clip=true,width=6.5in]{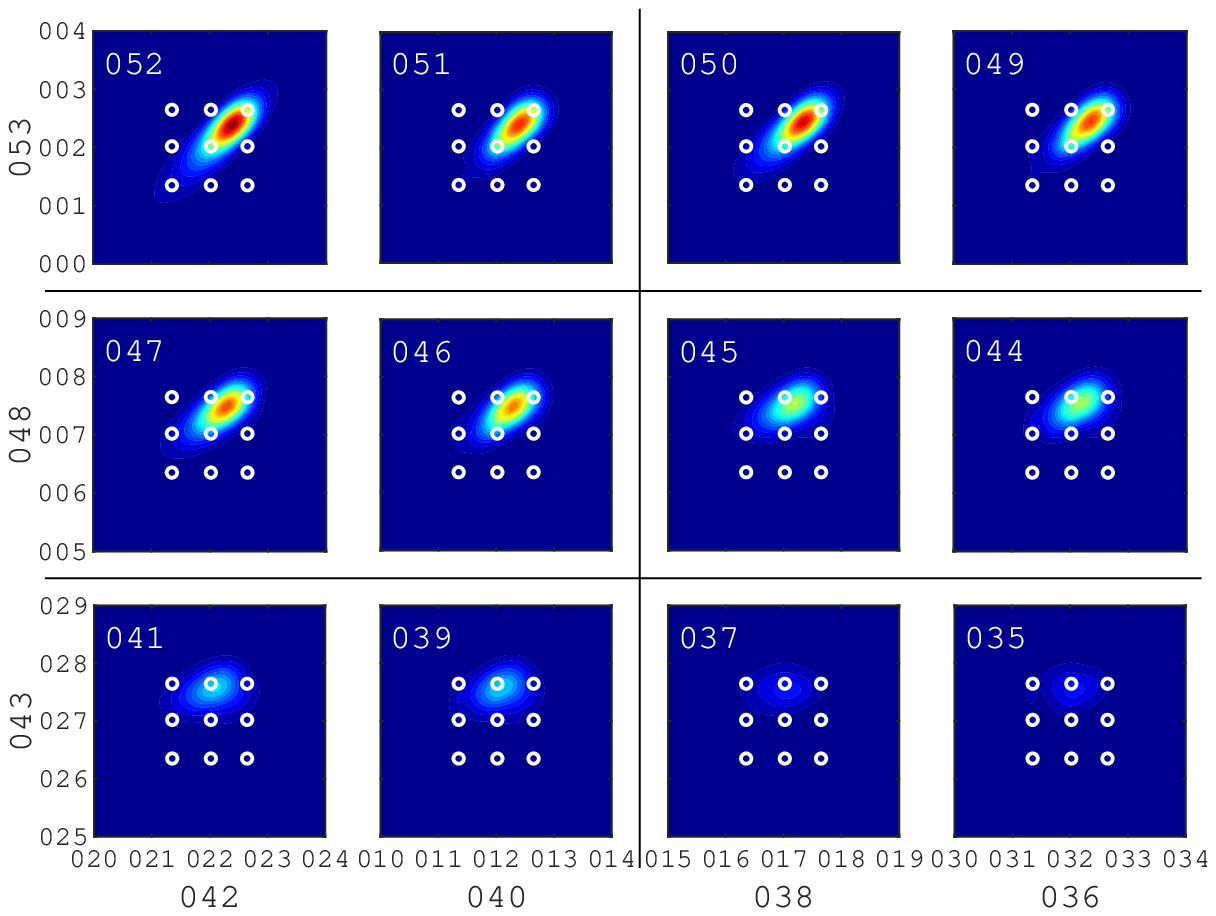}
\caption{Cross-correlation as a function of $x$ and $x^{\prime}$ for fixed time-lag values (a-b) $\tau = 0$; (c-d) $\tau = 1$; (e-f) $\tau = 2$; (g-h) $\tau = 4$; (i-j) $\tau = 6$; (k-l) $\tau = 10$.  In each case, the left-hand plot shows the true correlations, and the right-hand plot shows the estimated values using three probes, which lead to known cross-correlations at the locations indicated by the small circles. The contour levels are the same in each panel and range from zero to the maximum values of the true correlation at $\tau=0$.}
\label{fig:COR_estimate_3p}
\end{figure}

The space-time correlation tensor $\tc{C}_{\boldsymbol{q}\boldsymbol{q}}$ can be recovered from the cross-spectral density $\tc{S}_{\boldsymbol{q}\boldsymbol{q}}$ using the inverse Fourier transform of (\ref{Eq:COR2CSD}).  As an example, figure~\ref{fig:COR_estimate_Lines} shows the true (solid lines) and estimated (dashed lines) correlations as a function of time lag $\tau$ for three spatial locations, $x = -5$, 0 and 5. These locations correspond to a low-energy region, a probe position and the energy peak, respectively.  Each curve has been scaled by the maximum value of the corresponding true correlation.  

The one-point autocorrelation for each point is shown in figure~\ref{fig:COR_estimate_Lines}(a-c).  The amplitude of the autocorrelation for the low-energy point at $x=-5$ (panel a) is significantly under-predicted, but the correlation length scale is well captured.  The estimated autocorrelation at $x=0$ (panel b) is exact since this point corresponds to one of the probe locations.   The autocorrelation at $x=5$ (panel c) is accurately estimated apart from a small under-prediction of the peak, which corresponds to an under-prediction of the variance.  

The cross-correlations between these three points are shown in figure~\ref{fig:COR_estimate_Lines}(d-f).  The estimates are quite good in all cases, including those involving the low-energy point (panels d and f) and two unknown points (panel f).  It is interesting that the cross-correlations involving the low-energy point are more accurate than the autocorrelation at this point.  The agreement for the cross-correlation between the known and high-energy points (panel e) is almost perfect.  

The spatial distribution of the true and estimated cross-correlation tensors at fixed values of the time lag $\tau$ is shown in figure~\ref{fig:COR_estimate_3p}. The plotted time lag values range from $\tau = 0$ to 10; negative values need not be considered due to the symmetry 
\begin{equation}
\tc{C}_{\boldsymbol{q}\boldsymbol{q}}(\boldsymbol{x_{1}},\boldsymbol{x}_{2},-\tau) = \tc{C}_{\boldsymbol{q}\boldsymbol{q}}^{*}(\boldsymbol{x_{2}},\boldsymbol{x}_{1},\tau).
\end{equation}
The contour levels are the same in each panel and range from zero to the maximum values of the true correlation at $\tau=0$.  Again, the circles indicate the locations where the correlations are known, and the remaining values are to be estimated.

The spatial correlation tensor is obtained for $\tau = 0$ and is shown in figure~\ref{fig:COR_estimate_3p}(a).  As already observed in figure~\ref{fig:COR_estimate_Lines}, the amplitudes of the correlations at zero time lag are slightly underpredicted, but the spatial shape and overall amplitude are well captured.  As the time lag $\tau$ is increased, the estimates faithfully track the changing shape of the true correlations up to at least $\tau = 10$, by which point the magnitudes of the correlations are small.

\subsubsection{Impact of probe location and comparisons with the rank-1 model}
\label{Sec:GL_compare}

\begin{figure}
\input{PSD_estimate_Compare_v2_JFM.tex}
\centering
\includegraphics[trim=0cm 0.0625in 0cm 0.0cm, clip=true,width=6.5in]{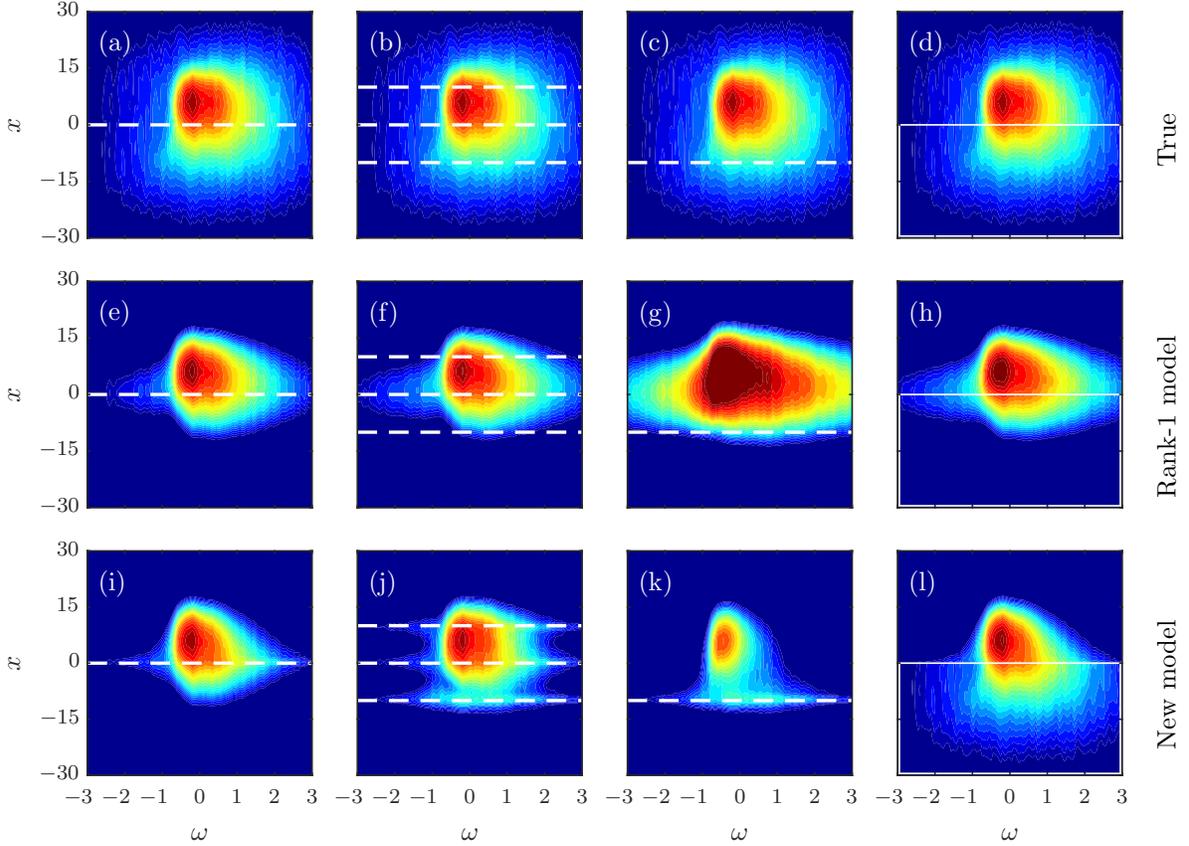}
\caption{Power spectral density as a function of $\omega$ and $x$: (a-d) true values repeated for ease of comparison; (e-h) estimated values from the rank-1 model of \cite{Beneddine:2016}; (i-l) estimated values from the new model presented in this paper.  The estimates are based on the following probe locations: (e,i) $x = 0$; (f,j) $x = -10,0$ and 10; (g,k) $x = -10$; (h,l) $x \leq 0$. The contour levels are the same as figure~\ref{fig:PSD_estimate_3p}. }
\label{fig:PSD_estimate_Compare}
\end{figure}

In this section, comparisons are made between the new method described in this paper and the rank-1 method of \cite{Beneddine:2016} that was discussed in \S~\ref{sec:intro}.  Particular attention is given to the impact of the probe location(s) on the accuracy of the estimates provided by these two methods.

The PSD, which is the target quantity of the rank-1 model, is considered first.  figure~\ref{fig:PSD_estimate_Compare} compares the true PSD (top row) to the estimates from the rank-1 method (second row) and the new model (third row) for four different sets of probe locations (columns).  

First, a single probe is places at $x = 0$.  In this case, the methods provide similar estimates, but the peak amplitude is slightly better predicted by the new method while higher frequencies are better captured by the rank-1 method.  Adding two more probes at $x = \pm 10$ (second column) has little impact on the rank-1 estimate.  In contrast, the new method is able to use this additional information to improve its estimate, particularly in regard to the shape of the moderately energetic region surrounding the peak and in the vicinity of the new probes.  

Next, a single probe is placed at the low-energy location $x = -10$, well away from the peak.  At this point, the underlying assumption of the rank-1 model --  that the solution is dominated by the leading resolvent mode -- is false.  This is representative of the situation that will be encountered in real turbulent flows.  Because of this, the rank-1 method leads to large over-predictions of the PSD.  In contrast, the new method yields a moderate under-prediction of the PSD, which can be attributed to neglecting the unobservable portions of $\tc{S}_{\boldsymbol{f}\boldsymbol{f}}$.  

In the final case (fourth column), the flow statistics are known in a continuous region, $x \leq 0$, rather than at an isolated set of points.  These results can be compared to the first case in which statistics were known only at the boundary of this region, $x=0$.  Including the additional data for $x < 0$ leads to worse results for the rank-1 model (notice the significant over-prediction of the peak).  This is an undesirable property; it means that poorly placed probes (where the rank-1 assumption is invalid) can obscure the information provided by well-placed probes (where the assumption is valid).  More generally, this is a manifestation of the fact that the rank-1 method does not necessarily converge with increasing input information, even in the limit of complete knowledge of the flow statistics \citep{Towne2018spectral}, in contrast to the new method.  In the current example, it is clear that the additional information for $x<0$ does not degrade the estimate of the new method as it did for the rank-1 model, but instead leads to small improvements in the estimated PSD.  

%In contrast, the new method converges by construction; adding additional probes always reduces the error in a global (in $x$) norm, and the error goes to zero in the limit of complete knowledge of the flow statistics

\begin{figure}
\input{CSD_estimate_Compare_v2_JFM.tex}
\centering
\includegraphics[trim=0cm 0.0625in 0cm 0.0cm, clip=true,width=6.5in]{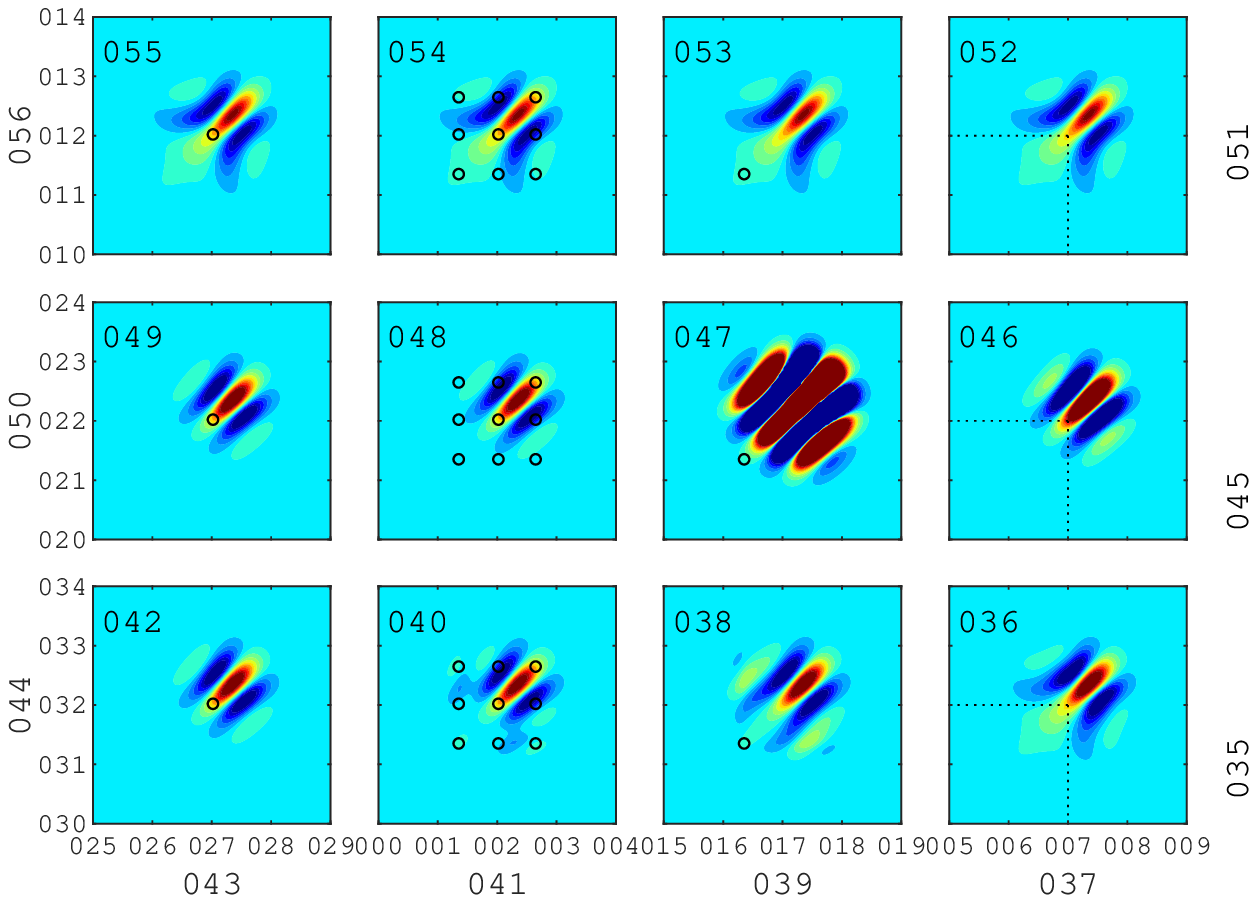}
\caption{Cross-spectral density as a function of $x$ and $x^{\prime}$ as the frequency of maximum gain, $\omega = -0.6$: (a-d) true values repeated for ease of comparison; (e-h) estimated values from the rank-1 model of \cite{Beneddine:2016}; (i-l) estimated values from the new model presented in this paper.  The estimates are based on the following probe locations: (e,i) $x = 0$; (f,j) $x = -10,0$ and 10; (g,k) $x = -10$; (h,l) $x \leq 0$.  The contour levels are the same in each panel and vary linearly between the minimum and maximum value of the true CSD.}
\label{fig:CSD_estimate_Compare}
\end{figure}

Next, comparisons are made between the the CSD estimates provided by the two methods for the same four sets of probes.  While the rank-1 method was not specifically designed to estimate cross-spectra, its form nevertheless implies values for these two-point statistics, and their accuracy is important if the method is to be used for time-domain modeling, as in \cite{Beneddine:2017}.  Comparisons are made for the frequency of maximum gain, $\omega = -0.6$, where the rank-1 model is expected to be most appropriate.  

The overall conclusions regarding the CSD estimates are similar to those just discussed for the PSD.  The two methods yield equivalent results for the single probe at $x = 0$.  The estimates for the new model are improved by adding two more probes at $x = \pm 10$, while they have little impact on the rank-1 results.  Using a single probe at the low-energy point $x = -10$ leads to reasonable estimates using the new method but large errors for the rank-1 method.  Finally, using data from the region $x \leq 0$ degrades the rank-1 estimates compared to using only the boundary point $x=0$ but improves the estimates obtained using the new model.  In this case, the CSD estimates from the new model are indistinguishable from the true CSD, even though the known region does not contain the peak PSD.

% -----------------------------------------------------------------------------
% -- SUBSECTION: Re180 Channel
% -----------------------------------------------------------------------------

\subsection{Turbulent channel flow} 
\label{Sec:Channel}

%=========================================================================
% -- Channel data --------------------------------------------------------
\subsubsection{Flow parameters, simulation and data processing}

\begin{figure}
\input{C180_mean_and_rms_JFM.tex}
\centering
\includegraphics[trim=0cm 0cm 0cm 0.0cm, clip=true,width=6.25in]{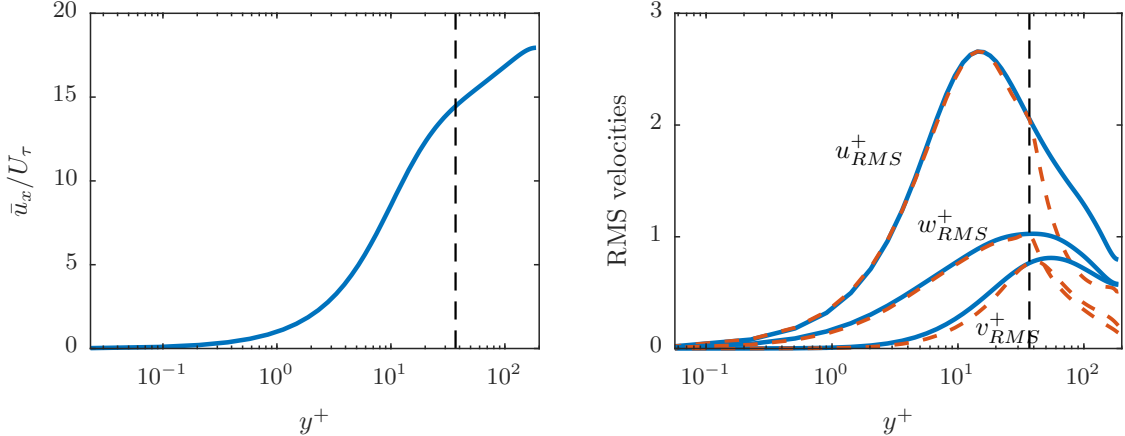}
\caption{ (a) Mean and (b) root-mean-squared velocities.  Solid lines: true values calculated from the DNS data.  Dashed lines: estimates obtained from the model using measurements at $y^{+} = 37$ ($y/h = 0.2$).  This input location is demarcated in the figure by the vertical dashed line.}
\label{fig:C180_rms}
\end{figure}

Next, we apply the new method to an incompressible turbulent channel flow at friction Reynolds number $Re_{\tau} = 187$, defined in terms of the friction velocity $U_\tau$ and the kinematic viscosity $\nu$. Wall units, denoted by $+$ superscripts, are also defined in terms of $U_\tau$ and $\nu$.  The flow is computed via direct numerical simulation (DNS) of the incompressible Navier-Stokes equations in a domain of size $x/h \times y/h \times z/h \in [0, 2 \pi] \times [0, 2] \times [0, \pi]$, where $x$, $y$, and $z$ are the streamwise, wall-normal, and spanwise dimensions and $h$ is the channel half-width.  The periodic directions $x$ and $z$ are discretized using 64 Fourier modes in each direction, and the wall-normal direction $y$ is discretized using 129 Chebyshev polynomials.  The flow is driven by impossing a constant mass flux in the streamwise direction.  The equations are advanced in time using a variable time step third-order Runge-Kutta integrator with a CFL number of 0.5.  To facilitate post processing, the data is interpolated in time to 10000 evenly spaced time instances with $\Delta t^{+} = 1.5$.  The mean streamwise velocity is shown in figure~\ref{fig:C180_rms}(a).

%\begin{figure}
%\input{../Figures/C180_meanU.tex}
%\centering
%\includegraphics[trim=0cm 0cm 0cm 0.0cm, clip=true,width=3.5in]{../Figures/C180_meanU.eps}
%\caption{Mean velocity profile.}
%\label{fig:C180_meanU}
%\end{figure}

The simulation data are used to compute the cross-spectral density tensor $\tc{S}_{\boldsymbol{q}\boldsymbol{q}}$, where $\boldsymbol{q} = \left[ u, v, w \right]^{T}$ and $u$, $v$, and $w$ are the streamwise, wall-normal, and spanwise velocities, respectively.  Since the flow is periodic in $x$ and $z$, the cross-spectral density is a function of wavenumber in these directions, i.e., $\tc{S}_{\boldsymbol{q}\boldsymbol{q}} = \tc{S}_{\boldsymbol{q}\boldsymbol{q}}(y,y^{\prime};k_{x}, k_{z}, \omega)$.  The cross-spectral density is estimated using Welch's (\citeyear{Welch:1967}) method.  The flow data are divided into overlapping blocks each containing $N_{fft}$ instantaneous snapshots of the flow.  A discrete Fourier transform in $x$, $z$, and $t$ is applied to each block, leading to Fourier modes of the form $\hat{\boldsymbol{q}}_{j}(y; k_{x}, k_{z}, \omega)$ for $j = 1,2,\dots,N_{b}$, where $N_{b}$ is the total number of blocks.  Then, the cross-spectral density is estimated as   
\begin{equation}
\label{Eq:Channel_CSD_estimate}
\tc{S}_{\boldsymbol{q}\boldsymbol{q}}(y,y^{\prime};k_{x}, k_{z}, \omega) = \frac{1}{N_{b}} \sum \limits_{j = 1}^{N_{b}} \hat{\boldsymbol{q}}_{j}(y; k_{x}, k_{z}, \omega) \hat{\boldsymbol{q}}_{j}^{*}(y^{\prime}; k_{x}, k_{z}, \omega).
\end{equation}
Finally, the estimated cross-spectra are further averaged according to the symmetries described by \cite{Sirovich:1987b}, which ensures that the estimated cross-spectra are symmetric with respect to reflection across the channel center line and to $180$ degree rotation about the $x$-axis.  We use blocks containing $N_{fft} = 256$ instantaneous snapshots with $75\%$ overlap, leading to $N_{b} = 156$ blocks, and we have verified that our results are insensitive to these choices.

%=========================================================================
% -- Channel LNS / resolvent ---------------------------------------------
\subsubsection{Linearized Navier-Stokes equations}

The resolvent operators required for the model are obtained from the incompressible Navier-Stokes equations

\begin{subequations}
\label{Eq:incomp_NavierStokes}
\begin{align}
\frac{\partial \boldsymbol{u}}{\partial t} + \bar{\boldsymbol{u}} \cdot \nabla \boldsymbol{u} +  \boldsymbol{u} \cdot \nabla \bar{\boldsymbol{u}} + \nabla p - \frac{1}{Re_{\tau}} \nabla \cdot \left[ \frac{\nu_{T}}{\nu} \left( \nabla \boldsymbol{u} + \nabla \boldsymbol{u}^{T} \right) \right] =& \, \boldsymbol{f}_{\boldsymbol{u}}, \\
\nabla \cdot \boldsymbol{u} =& \, 0,
\end{align}
\end{subequations}
where $\boldsymbol{u} = \left[ u, v, w \right]^{T}$ is a vector of velocity disturbances,  $\bar{\boldsymbol{u}} = \left[ \bar{u}, 0, 0 \right]$ is the mean velocity, and $p$ is the pressure disturbance.  
Following previous work \citep{reynolds_hussain_1972,delAlamo:2006,Illingworth:2018}, we have included an eddy viscosity model in the form of the total viscosity function $\nu_{T}(y)$.  Details of our formulation are consistent with those of \cite{Illingworth:2018} and can be found there.  

Since the linearized equations are homogeneous in $x$ and $z$, we can apply Fourier transforms in these directions and obtain an equation for each $(k_{x}, k_{z})$ wavenumber pair in the form of equation~(\ref{Eq:intro_model_LNS_time}) with 
\begin{equation}
\label{Eq:LNS_A_channel}
\mathcal{A} = i k_{x} A_{x} + A_{y} \frac{\partial}{\partial y} + i k_{z} A_{z} - k_{x}^{2} A_{xx} + A_{yy} \frac{\partial^{2}}{\partial y^{2}} - k_{z}^{2} A_{zz},
\end{equation}
$\Gamma = \mathrm{diag}\left( \left[ 1,1,1,0 \right] \right)$, and $\boldsymbol{q} = \left[ u, v, w, p\right]^{T}$.  The matrices in equation~(\ref{Eq:LNS_A_channel}) are provided in Appendix \ref{AppendixB}.  The wall-normal direction $y$ is discretized using 201 Chebyshev polynomials, and no-slip boundary conditions are applied at the walls.  

We choose the known quantity $\boldsymbol{y}$ to correspond to the three velocity components at $y/h = 0.2$, which in inner units corresponds to $y^{+} = 37$.  This is the same $y/h$ value considered by \cite{Illingworth:2018} in their recent Kalman filter study, although the $y^{+}$ value is different due to differing Reynolds numbers.  This location is relevant to the application of LES wall modeling, in which one would use data along such a surface to approximate the near-wall flow and/or shear stress.

To visualize the results, we will focus on the velocity energy spectra, which are obtained from the cross-spectral density tensor as
\begin{equation}
\label{Eq:energies}
\tc{E}_{\boldsymbol{q}\boldsymbol{q}}(y;k_{x}, k_{z}, \omega) = \tc{S}_{\boldsymbol{q}\boldsymbol{q}}(y,y; k_{x}, k_{z}, \omega).
\end{equation}

%=========================================================================
% -- Channel RMS results -------------------------------------------------
\subsubsection{Root-mean-squared velocities}

We begin by examining the root-mean-squared (RMS) velocity fluctuations, which are obtained by integrating $\tc{E}_{\boldsymbol{q}\boldsymbol{q}}(y;k_{x}, k_{z}, \omega)$ in $k_{x}$, $k_{z}$, and $\omega$ and taking the square root.  The true RMS velocity fluctuations computed from the DNS data and those obtained from the model are compared in figure~\ref{fig:C180_rms}(b) as a function of $y^{+}$.  The RMS values are accurately estimated for all three velocity components in the near-wall region, specifically for $y^{+} \lesssim 45$ ($y/h \lesssim 0.25$).  The streamwise velocity estimates are especially accurate, while slightly larger discrepancies are observed for the wall normal velocity.  Notably, the model accurately captures both the location and magnitude of the $u_{RMS}$ peak.  For larger values of $y^{+}$, the RMS values quickly fall below the DNS values.  Results in the following section show that this under prediction is due to missing energy at small scales which do not have a footprint at the probe location.  This missing energy could potentially be recovered by appropriate modeling of the $\boldsymbol{E}_{ij}$ terms that have been set to zero, as discussed in \S~\ref{Sec:conclusions}.

%=========================================================================
% -- Channel 1D spectra --------------------------------------------------
\subsubsection{Energy spectra}

\begin{figure}
\input{C180_kx_y_spectrum_log.tex}
\centering
\includegraphics[trim=0cm 0.2cm 0cm 0.0cm, clip=true,width=6.25in]{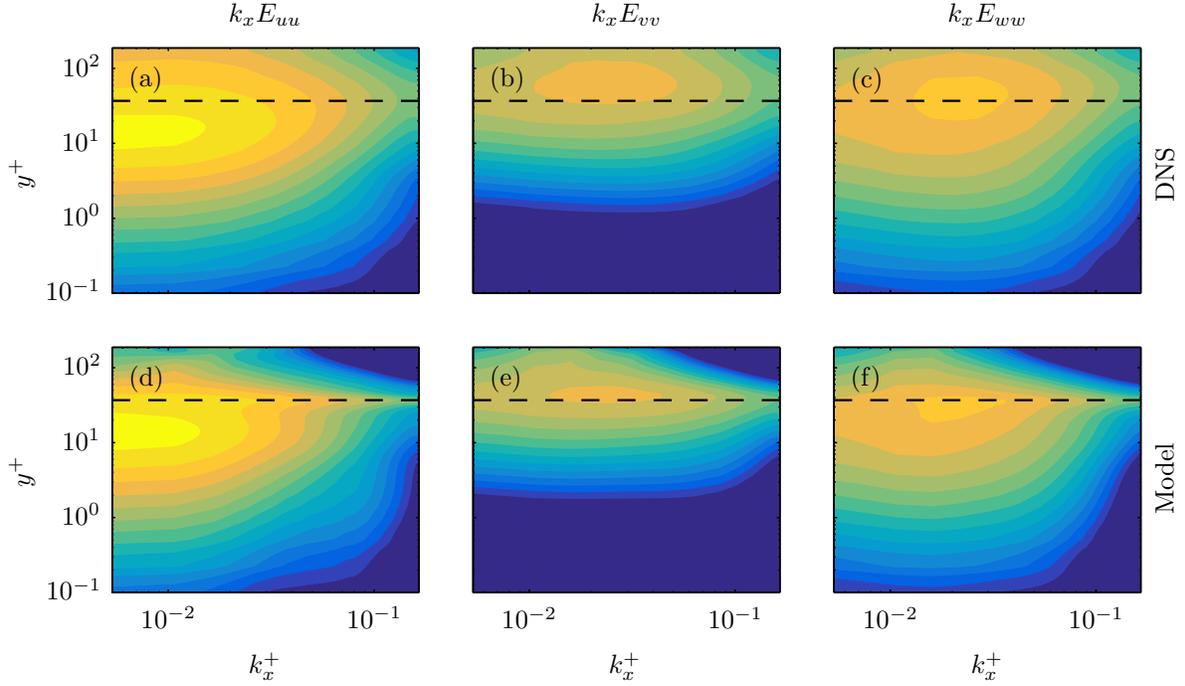}
\caption{Premultiplied energy spectra as a function of streamwise wavenumber $k_{x}^{+}$ and wall-normal distance $y^{+}$.  Top row: DNS.  Bottom row: estimates from the model.  Columns from left to right: streamwise velocity, wall-normal velocity, spanwise velocity.  The contour levels are logarithmically spaced and span five orders of magnitude, with the highest level equal to the maximum value of the DNS streamwise velocity spectrum.  The horizontal dashed lines show the location of the known input data, $y^{+} = 37$.}
\label{fig:C180_kx_y_spectrum}
\end{figure}

\begin{figure}
\input{C180_kz_y_spectrum_log.tex}
\centering
\includegraphics[trim=0cm 0.2cm 0cm 0.0cm, clip=true,width=6.25in]{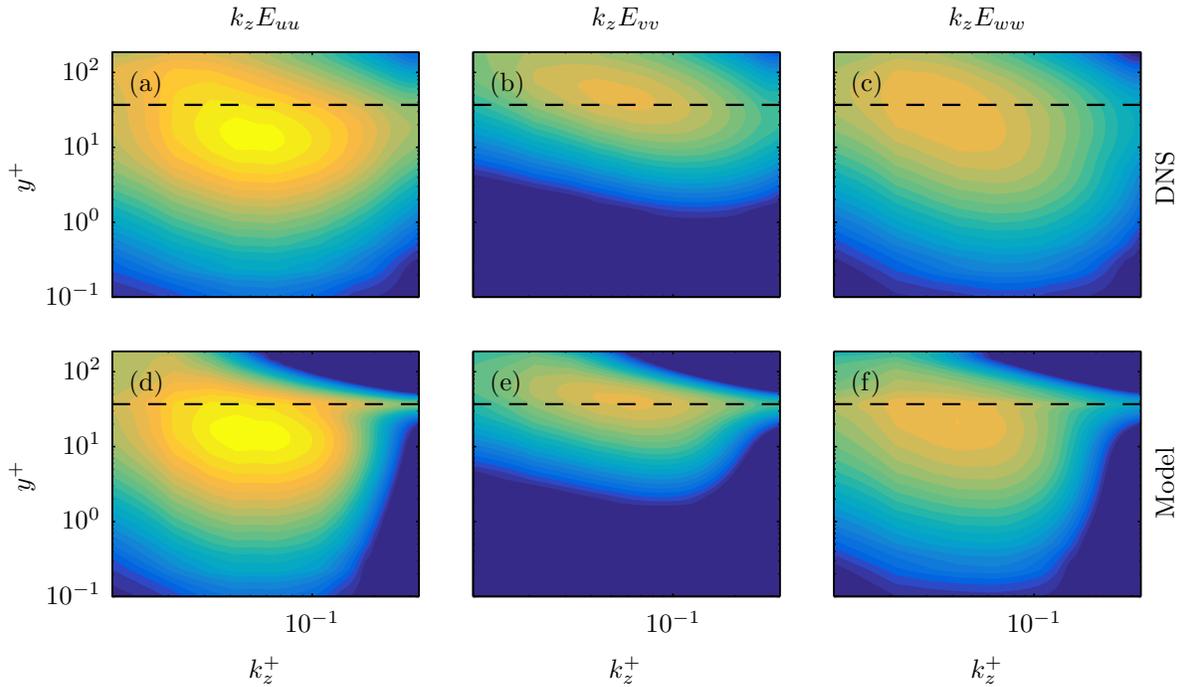}
\caption{Premultiplied energy spectra as a function of spanwise wavenumber $k_{z}^{+}$ and wall-normal distance $y^{+}$.  Details are the same as figure~\ref{fig:C180_kx_y_spectrum}.}
\label{fig:C180_kz_y_spectrum}
\end{figure}

\begin{figure}
\input{C180_omega_y_spectrum_log.tex}
\centering
\includegraphics[trim=0cm 0.2cm 0cm 0.0cm, clip=true,width=6.25in]{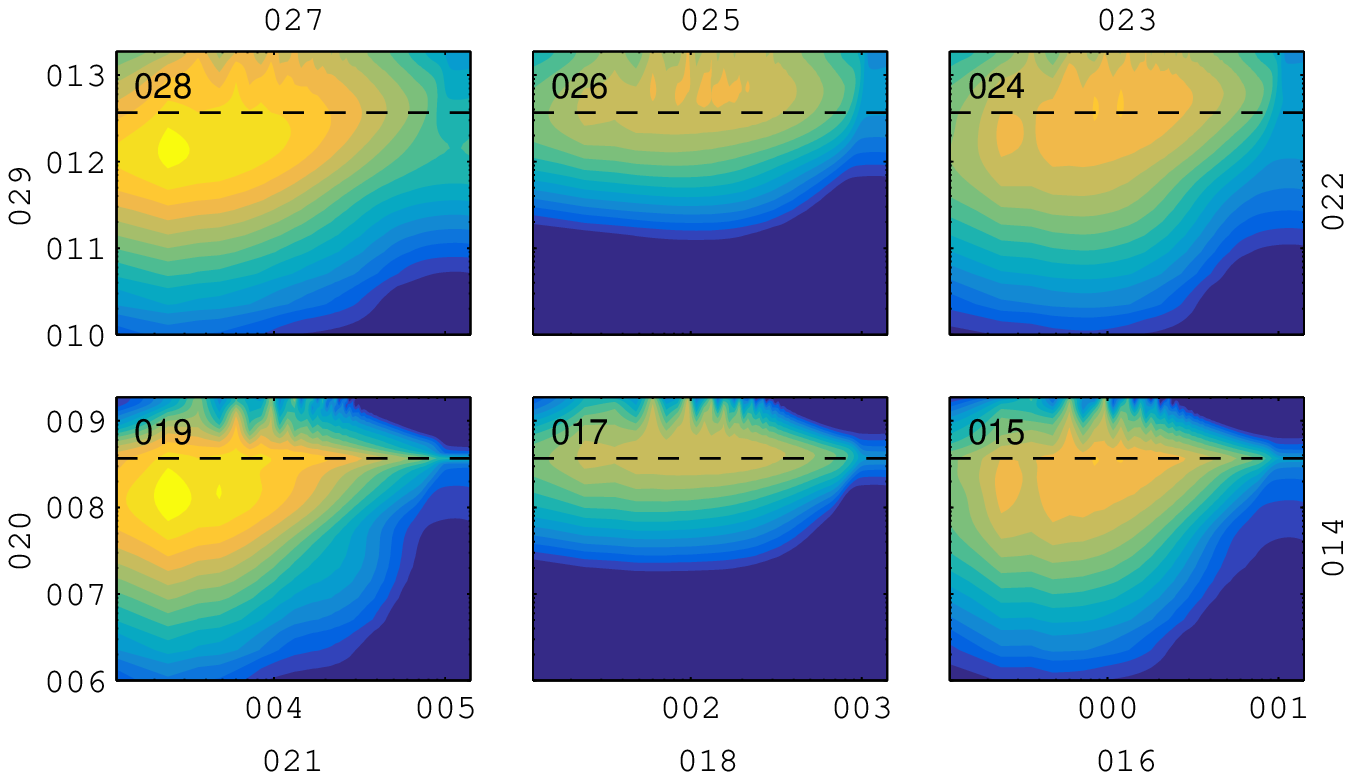}
\caption{Premultiplied energy spectra as a function of frequency $\omega^{+}$ and wall-normal distance $y^{+}$.  Details are the same as figure~\ref{fig:C180_kx_y_spectrum}.}
\label{fig:C180_omega_y_spectrum}
\end{figure}

Figures~\ref{fig:C180_kx_y_spectrum},~\ref{fig:C180_kz_y_spectrum}, and~\ref{fig:C180_omega_y_spectrum} show the energy spectrum for each velocity component as a function of $y^{+}$ and $k_{x}^{+}$, $k_{z}^{+}$, and $\omega^{+}$, respectively.  In each case, the energy has been integrated over the other two Fourier variables.  The energies have been premultiplied by the appropriate wavenumber or frequency to account for the logarithmic axes. The contour levels are logarithmically spaced and span five orders of magnitude, with the highest level equal to the maximum value of the DNS streamwise velocity spectrum.  The same levels are used in all subplots so that magnitudes can be directly compared. The true spectra computed from the DNS data appear in the top row of each figure, and the corresponding model estimates appear in the second row.

In all cases, the model accurately captures the energy distribution of all three velocity components for $y^{+} \lesssim 45$, except at the highest wavenumbers and frequencies. The amplitudes and locations of the energy peaks in $(y^{+}, k_{x}^{+}, k_{z}^{+}, \omega^{+})$ space are captured by the model.  On the other hand, the model under-predicts the energy at all wavenumbers and frequencies for higher values of $y^{+}$, which is consistent with the under-prediction of the RMS values observed in figure~\ref{fig:C180_rms}.  The highest wavenumbers and frequencies are correctly predicted only near the position of the known input data at $y^{+} = 37$ (horizontal dashed lines in the figures).

\begin{figure}[!h]
\input{C180_kx_omega_spectrum.tex}
\centering
\includegraphics[trim=0.0cm 0.5cm 0cm 0.1cm, clip=true,width=6.25in]{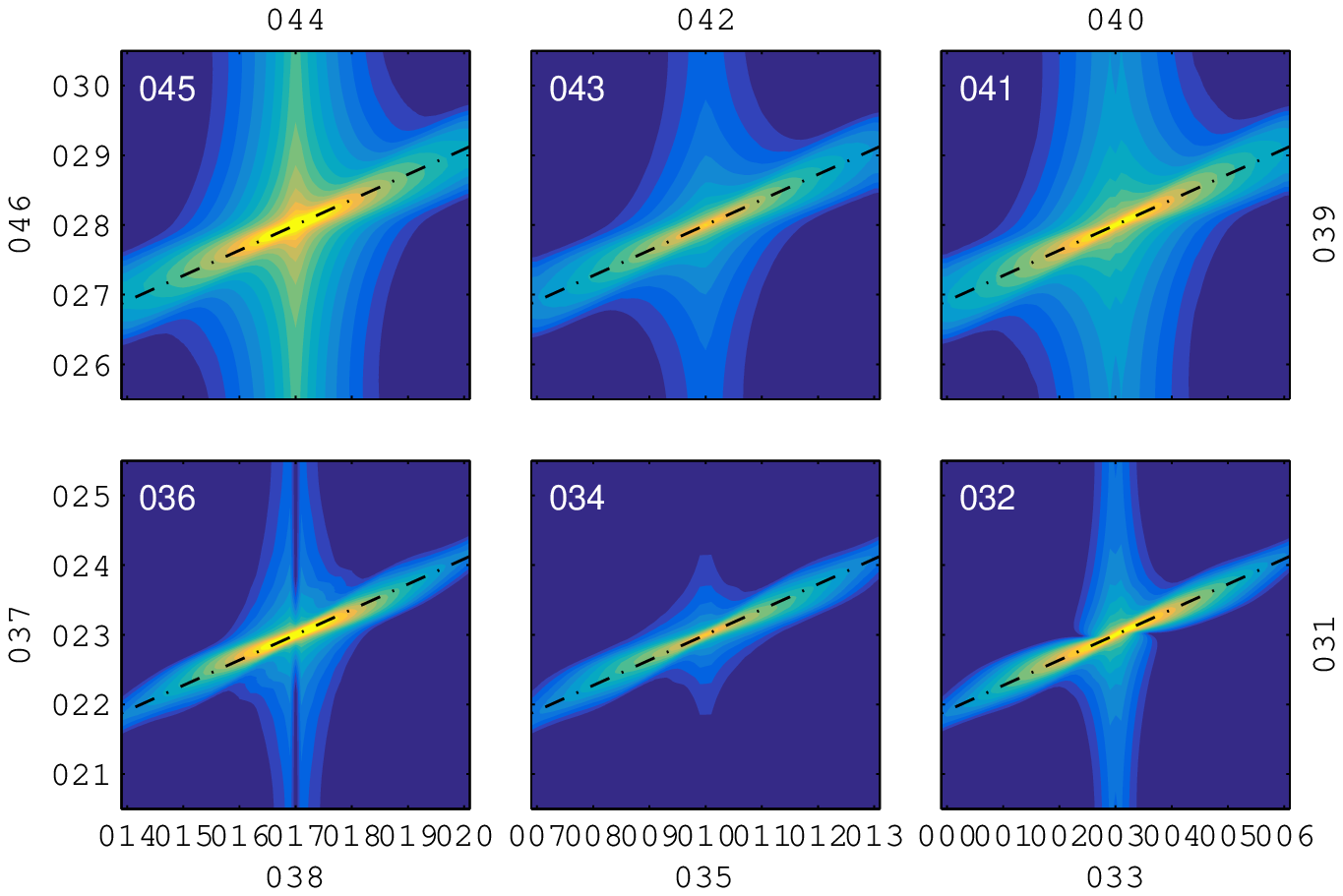}
\caption{Energy spectra as a function of streamwise wavenumber $k_{x}$ and frequency $\omega$ at $y^{+} = 9$ ($y/h = 0.05$).  Top row: DNS.  Bottom row: estimates obtained from the model.  Columns from left to right: streamwise, wall-normal, and spanwise velocity, respectively. The contour levels are logarithmically spaced and span five orders of magnitude, with the highest level equal to the maximum value of the DNS streamwise velocity spectrum. The dashed lines show the dominant phase speed at this wall-normal position.}
\label{fig:C180_kx_omega_spectrum}
\end{figure}

\begin{figure}[h!]
\input{C180_kx_omega_spectrum_Ycompare.tex}
\centering
\includegraphics[trim=0.0cm 0.5cm 0cm 0.1cm, clip=true,width=6.25in]{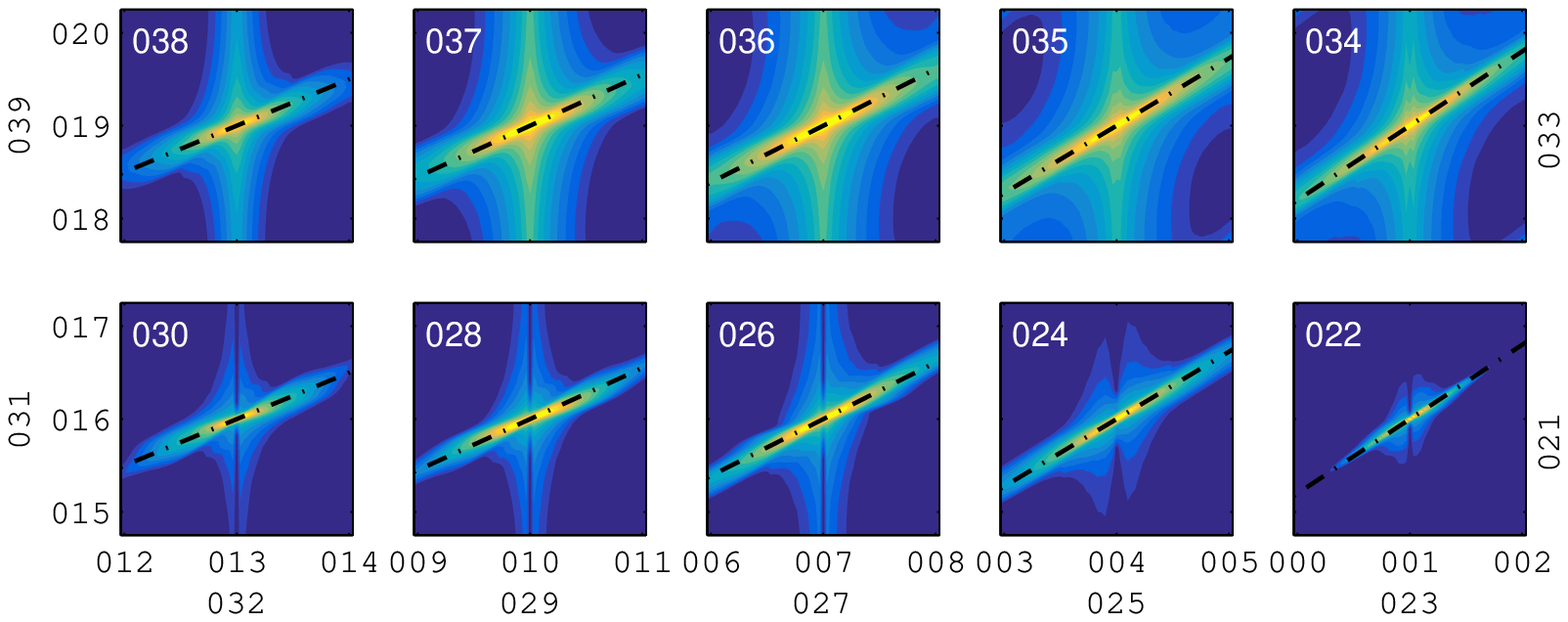}
\caption{Streamwise velocity energy spectra as a function of streamwise wavenumber $k_{x}$ and frequency $\omega$ at different wall-normal positions.  Top row: DNS.  Bottom row: estimates obtained from the model.  Columns from left to right: $y^{+} = 2,9,19,56,94$ ($y/h = 0.01, 0.05, 0.1, 0.3, 0.5$). The contour levels are logarithmically spaced and span five orders of magnitude, with the highest level equal to the maximum value of the DNS streamwise velocity spectrum at $y/h = 0.2$.  The dashed lines show the dominant phase speed at each wall-normal position.}
\label{fig:C180_kx_omega_spectrum_Ycompare}
\end{figure}

Figure~\ref{fig:C180_kx_omega_spectrum} shows the energy of each velocity component as a function of the streamwise wavenumber $k_{x}$ and frequency $\omega$ at the wall-normal position $y/h = 0.05$ ($y^{+} = 9$).  Here, we use linear axes so that the phase velocity $c_{p} = \omega / k_{x}$ can be easily visualized.  The contour levels are logarithmically spaced between the maximum value of the streamwise spectrum from DNS and span five orders of magnitude.  The same levels are used in all subplots.

Beginning with the DNS spectra shown in the top row, we see that the spectra are dominated by a band of energy that is approximately linear in $k_{x}-\omega$ space.  The slope of this line corresponds to the phase velocity  of the most energetic disturbances.  At this wall-normal location, the dominant phase velocity is $c_{p} / U_{\tau} \approx 11$, which is shown as a dashed line in each subplot.  The model accurately predicts the dominant phase velocity, and the main errors are observed primarily at phase velocities that are significantly different from the dominant one.

Figure~\ref{fig:C180_kx_omega_spectrum_Ycompare} shows the $k_{x} - \omega$ spectra for the streamwise velocity at five different wall-normal locations: $y^{+} = 2,9,19,56,94$ ($y/h = 0.01, 0.05, 0.1, 0.3, 0.5$).  It is clear that the model correctly captures the changes in phase velocity as a function of wall-normal position, even when the absolute energy levels are under predicted far from the wall.

%=========================================================================
% -- Channel cross-spectra -----------------------------------------
%\FloatBarrier
\subsubsection{Cross-spectra}

\begin{figure}
\input{C180_CSD_JFM.tex}
\centering
\includegraphics[trim=0cm 0.3cm 0cm 0.0cm, clip=true,width=6.25in]{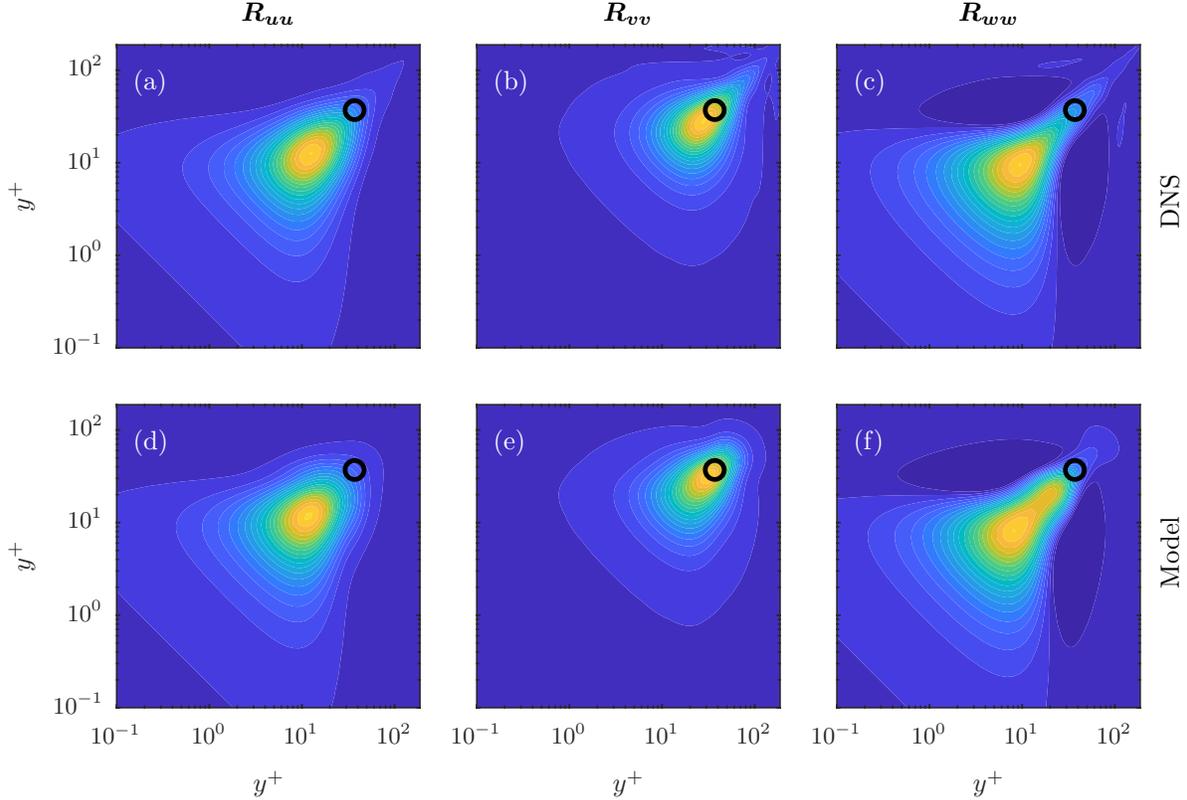}
\caption{Cross-spectral density for the spectral parameters $\lambda_{x}^{+} \approx 700$, $\lambda_{z}^{+} \approx 100$, and $c_{p} \approx 10$.  Top row: DNS.  Bottom row: estimates obtained from the model.  Columns from left to right: streamwise, wall-normal, and spanwise velocity, respectively. The black circle shows the input location where the DNS data is provided to the model.}
\label{fig:C180_CSD_JFM}
\end{figure}

In addition to the energy spectra considered so far, the model also provides predictions for cross-spectra.  An example is shown in figure~\ref{fig:C180_CSD_JFM}.  The CSD is plotted as a function of $y^{+}$ and $\left(y^{+}\right)^{\prime}$ for the the streamwise and spanwise wavenumbers $\lambda^{+}_{x} = 700$ and $\lambda^{+}_{z} = 100$, respectively, and the phase speed $c_{p} = 10$, which are typical values at which coherent structures are expected to appear \citep{Sharma:2013}.  The model uses the input data indicated by the black circles, and accurately reproduces the CSD for all three velocity components.

%=========================================================================
% -- Channel 1D autocorrelations -----------------------------------------

\subsubsection{Autocorrelations}

Next, we consider the space-time correlations
\begin{equation}
\label{Eq:spaceTime_corr_channel_def}
\tc{C}_{\boldsymbol{q}\boldsymbol{q}} \left( y,y^{\prime}, \delta x, \delta z, \delta t \right) = E \left\{ \boldsymbol{q}(x,y,z,t) \boldsymbol{q}^{*}(x+\delta x, y^{\prime}, z + \delta{z}, t + \delta t ) \right\},
\end{equation}
where the expectation is taken over all $x$, $z$, and $t$.  These correlations can be recovered from the cross-spectra discussed so far by taking inverse Fourier transforms,
\begin{equation}
\label{Eq:spaceTime_corr_channel_invFourier}
\tc{C}_{\boldsymbol{q}\boldsymbol{q}}(y,y^{\prime},\delta x, \delta z, \delta t) = \frac{1}{(2\pi)^{3}} \int \limits_{-\infty}^{\infty} \int \limits_{-\infty}^{\infty} \int \limits_{-\infty}^{\infty} \tc{S}_{\boldsymbol{q}\boldsymbol{q}}(y,y^{\prime},k_{x}, k_{z}, \omega) e^{-i k_{x} \delta x}e^{-i k_{z} \delta z}e^{-i \omega \delta t} d k_{x} d k_{z} d \omega.
\end{equation}
We will focus on the autocorrelations
\begin{equation}
\label{Eq:spaceTime_corr_channel_autocorr}
\boldsymbol{R}_{\boldsymbol{q}\boldsymbol{q}}\left(y; \delta x, \delta z, \delta t \right) = \tc{C}_{\boldsymbol{q}\boldsymbol{q}} \left( y,y, \delta x, \delta z, \delta t \right).
\end{equation}

\begin{figure}[h!]
\input{C180_x_t_correlation.tex}
\centering
\includegraphics[trim=0cm 0.5cm 0cm 0.0cm, clip=true,width=6.25in]{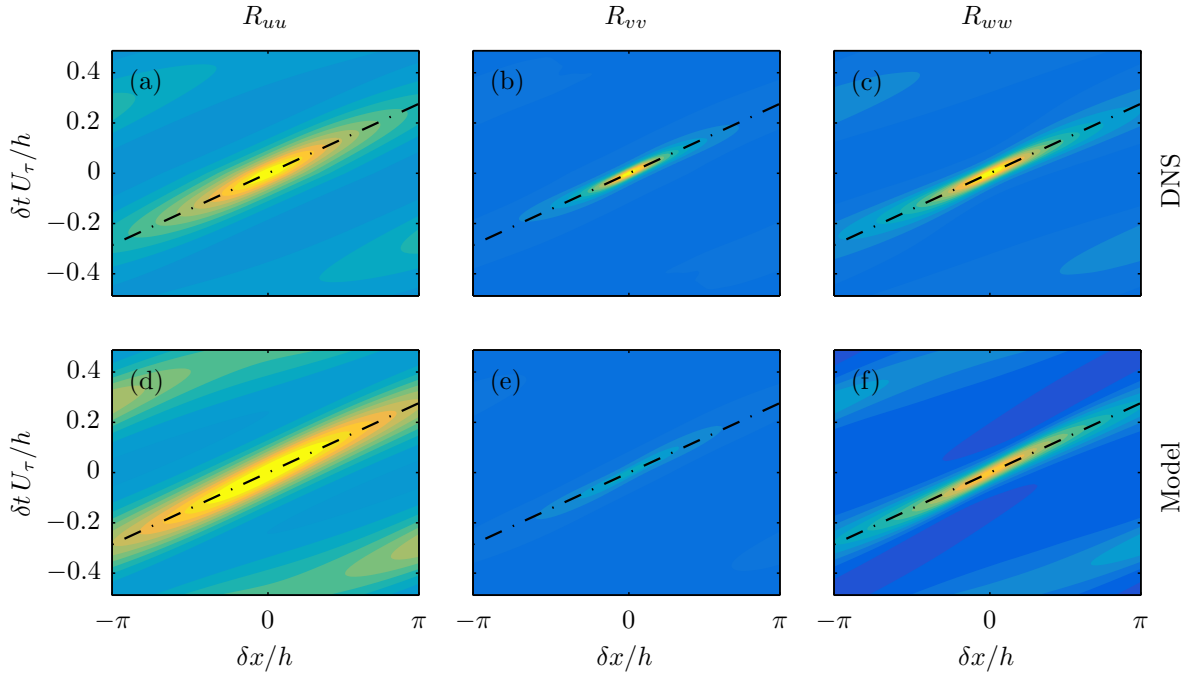}
\caption{Autocorrelation as a function of streamwise separation $\delta x$ and temporal separation $\delta t$ at $y^{+} = 9$ ($y/h = 0.05$).  Top row: DNS.  Bottom row: estimates obtained from the model.  Columns from left to right: streamwise, wall-normal, and spanwise velocity, respectively.  The contour levels are linearly spaced between $90\%$ and $-20\%$ of the maximum value of the DNS autocorrelation of each velocity component.  The slope of the dashed lines give the inverse of the dominant convection velocity.}
\label{fig:C180_x_t_correlation}
\end{figure}

\clearpage
As an example, we examine the autocorrelations as a function of the streamwise and temporal lag variables $\delta x$ and $\delta t$, respectively, at a fixed wall-normal location $y/h = 0.05$ ($y^{+} = 9$).  Figure~\ref{fig:C180_x_t_correlation} shows the autocorrelation of each velocity component as a function of $\delta x$ and $\delta t$, i.e., the space-time autocorrelations along the streamwise direction.  The contour levels are defined in the same way as in the previous three figures.  The inverse slope of the band of high correlation in each plot provides a measure of the convection velocity of disturbances.  At this wall-normal location, the convection velocity is approximately  $11 U_{\tau}$, which is consistent with the phase velocity shown in figure~\ref{fig:C180_kx_omega_spectrum} as well as the observations of \cite{Kim:1993}.  The convection velocity is accurately approximated by the model for all three velocity components.  The correlation magnitudes are also well approximated aside from a moderate under prediction of the peak wall-normal velocity correlations.  

%Figure~\ref{fig:C180_x_z_correlation} shows the autocorrelations as a function of $\delta x$ and $\delta z$ for $y^{+} = 9$ ($y/h = 0.05$).  The model provides accurate predictions for all three velocity components.  

%%%%%%%%%%%%%%%%%%%%%%%%%%%%%%%%%%%%%%%%%%%%%%%%%%%%%%%%%%%%%%%%%%%%%%%%%%%%%%%
% -----------------------------------------------------------------------------
% -- SECTION:
% -----------------------------------------------------------------------------

\section{Conclusions}
\label{Sec:conclusions}

Building on the work of \cite{Beneddine:2016} and \cite{Zare:2017}, this paper introduces a method for estimating space-time flow statistics from a limited set of known values.  The method is based on the resolvent methodology developed my \cite{Mckeon:2010} and the statisitical interpretation of this theory proposed by \cite{Towne2018spectral}.  The central idea of our approach is to (partially) infer the non-linear term of the linearized Navier-Stokes equations using limited flow data. This non-linear term is then used as a forcing acting on the resolvent operator to reconstruct unknown statistics of the flow. 
%The central idea of our approach is to use known data to infer the statistics of the nonlinear terms that constitute a forcing on the linearized Navier-Stokes equations.  These forcing statistics then imply values for the remaining unknown flow statistics through application of the resolvent operator.

The method has been demonstrated using two examples problems.  First, we applied it to the complex Ginzburg-Landau equation, which serves as a convient model of a convectively unstable flow.  Using input data from three probe locations, the method provides good estimates of the unknown power spectra, cross-spectra, and space-time correlations within the energetic regions of $\omega$-$x$ or $\tau$-$x$ space.  Comparisons are then made with the rank-1 model proposed by \cite{Beneddine:2016}.  The two methods give similar results when the probes are placed at locations dominated by a single resolvent mode, but the new method gives superior results when the probes are placed at locations that violate this underlying assumption of the rank-1 model.  The improved behavior in this case is important for turbulent flows, which cannot in general be described by a single resolvent mode.  Furthermore, the estimates provided by the new method improve with the addition of more known input data.

Second, we applied our method to a turbulent channel at friction Reynolds number $Re_{\tau} = 187$.  Using data exclusively from the wall-normal location $y/h = 0.2$ ($y^{+} = 37$), the method provides good estimates of the velocity energy spectra and autocorrelations for $y \lesssim 0.25$ ($y^{+} \lesssim 45$).  The energies and autocorrelations are under-predicted further away from the wall due to missing energy at small scales that requires additional modeling to capture.  The success of the model in the near-wall region using knowledge of the interior flow suggests that it could be useful for designing new wall models for large-eddy simulation that are capable of capturing fluctuations of wall quantities such as shear stress and heat transfer and near-wall velocities that play an important role, for example, in particle laden flows.

Additional work is required to understand these observations, further assess the impact of the location of the known input data, and determine whether the results described in this paper will extend to higher Reynolds numbers and other types of turbulent flows.  The properties and performance of the method should also be directly compared to other approaches that use the linearized flow equations as the basis for flow estimation, including the recent Kalman-filter-based approach described by \cite{Illingworth:2018}.  

The method itself could also be further improved by modeling the portions of the forcing cross-spectral density that can not be observed using the known data.  In the current formulation, these terms are simply set to zero, and there exist several possible alternatives.  One is to assume that the unobserved forcing is uncorrelated with the observed part and with itself, leading to the approximation
\begin{equation}
\tc{S}_{\boldsymbol{f}\boldsymbol{f}} = \left[ \boldsymbol{V}_{1} \,\,\,\boldsymbol{V}_{2} \right]     \left[ \begin{array}{cc} \boldsymbol{E}_{11} & 0 \\ 0 & a\boldsymbol{I} \end{array} \right]\left[ \boldsymbol{V}_{1} \,\,\,\boldsymbol{V}_{2} \right]^{*}.
\end{equation}
An appropriate value for the scalar amplitude $a$ could be determined from the amplitudes of the known $\tc{E}_{11}$ terms.  

Another possibility is to choose the unobservable terms by insisting that the estimated $\tc{S}_{\boldsymbol{f}\boldsymbol{f}}$ projects exclusively onto the first $n$ singular modes of $\mathcal{R}_{q}$. This possibility is similar to a suggestion made by \cite{Beneddine:2017}, except here the expansion coefficients are treated as statistical quantities rather than complex scalars.  As shown by \cite{Towne2018spectral}, this statistical treatment removes a fundamental accuracy restriction imposed by treating the expansion coefficients as deterministic scalars and allows for a convergent approximation.

%%%%%%%%%%%%%%%%%%%%%%%%%%%%%%%%%%%%%%%%%%%%%%%%%%%%%%%%%%%%%%%%%%%%%%%%%%%%%%%
% -----------------------------------------------------------------------------
% -- Backmatter 
% -----------------------------------------------------------------------------

\section*{Acknowledgments}

A.T. thanks Eduardo Martini, Andr\'e Cavalieri and Peter Jordan for fruitful discussion on this work.  A.L.D. was partially supported by NASA under Grant NNX15AU93A and by ONR under Grant N00014-16-S-BA10.

% ----------------------------------------------------------------------------
% --------------- APPENDICES
% ----------------------------------------------------------------------------

%\appendix

\begin{appendices}

% -----------------------------------------------------------------------------
% -- APPENDIX:
% -----------------------------------------------------------------------------

\section{Least-squares approximation of forcing}
\label{Sec:app_LS_forcing}

Let us compute the approximation of $\boldsymbol{S}_{\boldsymbol{f}\boldsymbol{f}}$ we would obtain using the pseudo-inverse of $\mathcal{R}_{{y}}$, which can be written in terms of the SVD (\ref{Eq:Ry_SVD}) as 
\begin{equation}
\label{Eq:Ry_pseudo_inv}
\mathcal{R}_{{y}}^{+} = \boldsymbol{V}_{\boldsymbol{y}} \boldsymbol{\Sigma}_{\boldsymbol{y}}^{-1} \boldsymbol{U}_{\boldsymbol{y}}^{*},
\end{equation}
where $\boldsymbol{\Sigma}_{\boldsymbol{y}}^{-1} = \left[ \boldsymbol{\Sigma}_{1}^{-1} \,\,\, \boldsymbol{0} \right]^{T}$.  Applying the pseudo-inverse and its complex conjugate to the left- and right-hand-sides of (\ref{Eq:Syy_res}), respectively, gives the pseudo-inverse approximation of $\boldsymbol{S}_{\boldsymbol{f}\boldsymbol{f}}$:
\begin{subequations}
\label{Eq:Sff_LS}
\begin{align}
\tc{S}_{\boldsymbol{f}\boldsymbol{f}}^{LS} =& \, \mathcal{R}_{y}^{+} \tc{S}_{\boldsymbol{y}\boldsymbol{y}} (\boldsymbol{R}_{y}^{+})^{*} \label{Eq:Sff_LS_1} \\
=& \, \boldsymbol{V}_{\boldsymbol{y}} \boldsymbol{\Sigma}_{\boldsymbol{y}}^{-1} \boldsymbol{U}_{\boldsymbol{y}}^{*} \tc{S}_{\boldsymbol{y}\boldsymbol{y}} \boldsymbol{U}_{\boldsymbol{y}} \boldsymbol{\Sigma}_{\boldsymbol{y}}^{-1} \boldsymbol{V}_{\boldsymbol{y}}^{*} \label{Eq:Sff_LS_2} \\
=& \, \left[ \boldsymbol{V}_{1} \,\,\,\boldsymbol{V}_{2} \right] \left[ \begin{array}{cc} \boldsymbol{\Sigma}_{1}^{-1} & \boldsymbol{0}\end{array} \right]^{T} \boldsymbol{U}_{\boldsymbol{y}}^{*} \tc{S}_{\boldsymbol{y}\boldsymbol{y}} \boldsymbol{U}_{\boldsymbol{y}} \left[ \begin{array}{cc} \boldsymbol{\Sigma}_{1}^{-1} & \boldsymbol{0}\end{array} \right] \left[ \boldsymbol{V}_{1} \,\,\,\boldsymbol{V}_{2} \right]^{*} \label{Eq:Sff_LS_3} \\
=& \, \boldsymbol{V}_{1} \boldsymbol{\Sigma}_{1}^{-1} \boldsymbol{U}_{\boldsymbol{y}}^{*} \tc{S}_{\boldsymbol{y}\boldsymbol{y}} \boldsymbol{U}_{\boldsymbol{y}} \boldsymbol{\Sigma}_{1}^{-1} \boldsymbol{V}_{1}^{*} \label{Eq:Sff_LS_4} \\
=& \, \boldsymbol{V}_{1} \boldsymbol{E}_{11} \boldsymbol{V}_{1}^{*} \label{Eq:Sff_LS_5}
\end{align}
\end{subequations}
with $\boldsymbol{E}_{11}$ given by (\ref{Eq:E11_eq}).  Equation~(\ref{Eq:Sff_LS_5}) is identical to~(\ref{Eq:Sff_approx}), which shows that setting the unknown $\boldsymbol{E}_{ij}$ terms to zero is equivalent to a least-squares, pseudo-inverse approximation of $\tc{S}_{\boldsymbol{f}\boldsymbol{f}}$.

% -----------------------------------------------------------------------------
% -- APPENDIX:
% -----------------------------------------------------------------------------

\section{Linearized incompressible Navier-Stokes operators}
\label{AppendixB}

The matrices defining the linearized incompressible Navier-Stokes equations in equations~(\ref{Eq:intro_model_LNS_time}) and~(\ref{Eq:LNS_A_channel}) are:

\begin{equation*}
\tc{A}_{x} =\left[\begin{array}{cccc} \bar{u} & -\nu_{T}^{\prime} & 0 & 1 \\ 0 & \bar{u} & 0 & 0 \\ 0 & 0 & \bar{u} & 0 \\ 1 & 0 & 0 & \bar{u}  \end{array}\right], \quad \tc{A}_{y} =\left[\begin{array}{cccc} -\nu_{T}^{\prime} & 0 & 0 & 0 \\ 0 & -2\nu_{T}^{\prime} & 0 & 1 \\ 0 & 0 & -\nu_{T}^{\prime} & 0 \\ 0 & 1 & 0 & 0 \end{array}\right], \quad \tc{A}_{z} =\left[\begin{array}{cccc} 0 & 0 & 0 & 0 \\ 0 & 0 & 0 & 0 \\ 0 & 0 & 0 & 1 \\ 0 & -\nu_{T}^{\prime} & 1 & 0  \end{array}\right],
\end{equation*}

\begin{equation*}
\tc{A}_{0} =\left[\begin{array}{cccc} 0 & \frac{\partial \bar{u}}{\partial y} & 0 & 0 \\ 0 & 0 & 0 & 0 \\ 0 & 0 & 0 & 0 \\ 0 & 0 & 0 & 0 \end{array}\right], \quad \tc{\Gamma} = \left[\begin{array}{cccc} 1 & 0 & 0 & 0 \\ 0 & 1 & 0 & 0 \\ 0 & 0 & 1 & 0 \\ 0 & 0 & 0 & 0 \end{array}\right], \quad \tc{B} = \left[\begin{array}{ccc} 1 & 0 & 0 \\ 0 & 1 & 0 \\ 0 & 0 & 1 \\ 0 & 0 & 0 \end{array}\right],
\end{equation*}
\vspace{0.1 in} and $\tc{A}_{xx} = \tc{A}_{yy} = \tc{A}_{zz} = -\frac{1}{Re} \tc{\Gamma}$.  We have defined $\nu_{T}^{\prime} = \frac{1}{Re} \frac{\partial \nu_T}{\partial y}$.

\end{appendices}

%%%%%%%%%%%%%%%%%%%%%%%%%%%%%%%%%%%%%%%%%%%%%%%%%%%%%%%%%%%%%%%%%%%%%%%%%%%%%

% Bibliography
\bibliographystyle{agsm}
%\bibliography{Bib_ResEst}

\end{document}